\documentclass[acmsmall,nonacm,screen]{acmart}

\newif\ifDraft  

\usepackage[utf8]{inputenc}

\usepackage{booktabs}
\usepackage{makecell}
\usepackage{pifont}
\newcommand{\y}{\ding{51}}
\newcommand{\n}{\ding{55}}
\newcommand{\fn}[1]{\textsuperscript{#1}}

\usepackage{arydshln}
\newcommand*\dhline{\specialrule{0pt}{1pt}{0pt}\hdashline[.4pt/3pt]\specialrule{0pt}{0pt}{1pt}}

\newcommand{\KFe}[1]{{\bf KF{#1}}}
\newcommand{\KF}[1]{(\KFe{#1})}

\ifDraft 
  \usepackage[textsize=small, textwidth=40mm]{todonotes}
\else 
  \usepackage[disable]{todonotes}
\fi

\AtBeginDocument{%
  }

\setcopyright{acmlicensed}
\copyrightyear{2026}
\acmYear{2026}
\acmDOI{XXXXXXX.XXXXXXX}

\begin{document}

\title[Digital Euro FAQs Revisited]{Digital Euro: Frequently Asked Questions Revisited}

\author{Joe Cannataci}
\affiliation{%
  \institution{University of Groningen}
  \city{Groningen}
  \country{Netherlands}
}
\orcid{0000-0001-6309-0734}
\email{j.a.cannataci@step-rug.nl}

\author{Benjamin Fehrensen}
\affiliation{%
  \institution{Bern University of Applied Science (BFH)}
  \city{Biel/Bienne}
  \country{Switzerland}
}
\orcid{0000-0002-7084-5434}
\email{benjamin.fehrensen@bfh.ch}

\author{Mikolai Gütschow}
\affiliation{%
  \institution{Dresden University of Technology (TUD)}
  \city{Dresden}
  \country{Germany}
}
\orcid{0009-0006-4489-7069}
\authornote{Corresponding author}
\email{mikolai.guetschow@tu-dresden.de}

\author{Özgür Kesim}
\affiliation{%
  \institution{Freie Universität Berlin}
  \city{Berlin}
  \country{Germany}
}
\orcid{0009-0003-7230-8282}
\email{o.kesim@fu-berlin.de}

\author{Bernd Lucke}
\affiliation{%
  \institution{University of Hamburg}
  \city{Hamburg}
  \country{Germany}
}
\orcid{0000-0002-4883-8756}
\email{bernd.lucke@uni-hamburg.de}

\renewcommand{\shortauthors}{Gütschow et al.}

\newcommand{\Privacy}{1}
\newcommand{\Offline}{2}
\newcommand{\Legal}{3}
\newcommand{\Economic}{4}
\newcommand{\Benefits}{5}
\newcommand{\Process}{6}

\begin{abstract}
The European Central Bank~(ECB) is working on the ``digital euro'',
	an envisioned retail central bank digital currency for the Euro area.
In this article, we take a closer look at the ``digital euro FAQ'',
	which provides answers to 27 frequently asked questions about the digital euro,
    and other published documents by the ECB on the topic.
We question the provided answers
	based on our analysis of the current design
	in terms of privacy, technical feasibility, risks, costs and utility.
In particular, we discuss the following key findings:
\begin{itemize}
    \item[\KF{\Privacy}] Central monitoring of all online digital euro transactions by the ECB
        threatens privacy even more than contemporary digital payment methods with segregated account databases.
    \item[\KF{\Offline}] The ECB's envisioned concept of a secure offline version of the digital euro offering full anonymity
		is in strong conflict with the actual history of hardware security breaches and mathematical evidence against it.
    \item[\KF{\Legal}] The legal and financial liabilities for the various parties involved remain unclear.
    \item[\KF{\Economic}] The design lacks well-specified economic incentives for operators
	as well as a discussion of its economic impact on merchants.
    \item[\KF{\Benefits}] The ECB fails to identify tangible benefits the digital euro would create for society,
	in particular given that the online component of the proposed infrastructure mainly duplicates existing payment systems.
	\item[\KF{\Process}] The design process has been exclusionary,
	with critical decisions being set in stone before public consultations.
	Alternative and open design ideas have not even been discussed by the ECB.
\end{itemize}
\end{abstract}

\ccsdesc[500]{Social and professional topics~Government technology policy}
\ccsdesc[500]{Information systems~Digital cash}
\ccsdesc[100]{Security and privacy~Tamper-proof and tamper-resistant designs}
\ccsdesc[300]{Social and professional topics~Surveillance}

\keywords{digital payment, security, retail CBDC, privacy of means of payment}

\received{\today}
\received[revised]{-}
\received[accepted]{-}

\maketitle

JEL: E42 · E58 · E52

\section{Introduction}

The Eurosystem under the leadership of the European Central Bank~(ECB)
	has been working for several years
	towards providing a retail central bank digital currency~(CBDC).
Since the presentation of the ECB's current plans
	(in the following referred to as ``the digital euro'')
	and a European Commission draft legislative proposal in summer 2023~\cite{ecDE2023},
	the ECB has started to provide information material to the public.
Among others, a list of frequently asked questions (FAQ) about the digital euro~\cite{defaq2025}
	which aim to address public concerns
    and to provide clarity on the proposed design and implications of the digital euro.
However, a deeper look reveals significant shortcomings in addressing key issues.
While the FAQ attempts to justify the digital euro,
    its current design raises more questions than it answers, particularly regarding privacy, security, economic effects, and overall utility.
The project risks being overly complex, economically burdensome, and technically flawed without clear benefits to users or society.

In this article, we provide a detailed criticism of the digital euro as currently proposed by the ECB.
\autoref{sec:design} starts with a brief overview of its design
	according to publicly available documents~\cite{ecDE2023, ecbFinalInv2023, ecbRulebookUpdate, ecbProgPrepFirst2024, ecbProgPrep2025}.
The subsequent sections \ref{sec:privacy}---\ref{sec:process}
    each give a summary and short explanation of our key findings and the major problems with this design.
Our detailed analysis of the answers of the ECB can be found in \autoref{sec:appendix:faq}---%
	clearly highlighting that the current answers fall short.
We conclude with recommendations for a successful digital euro in \autoref{sec:recommendations}.
The aim of this article is to provide a more comprehensive discussion of the current design to enable informed decisions,
	but not to argue against the introduction of any potential common digital payment system for the euro area.

\section{Digital Euro Design}
\label{sec:design}

\begin{figure}
	\includegraphics[width=\linewidth]{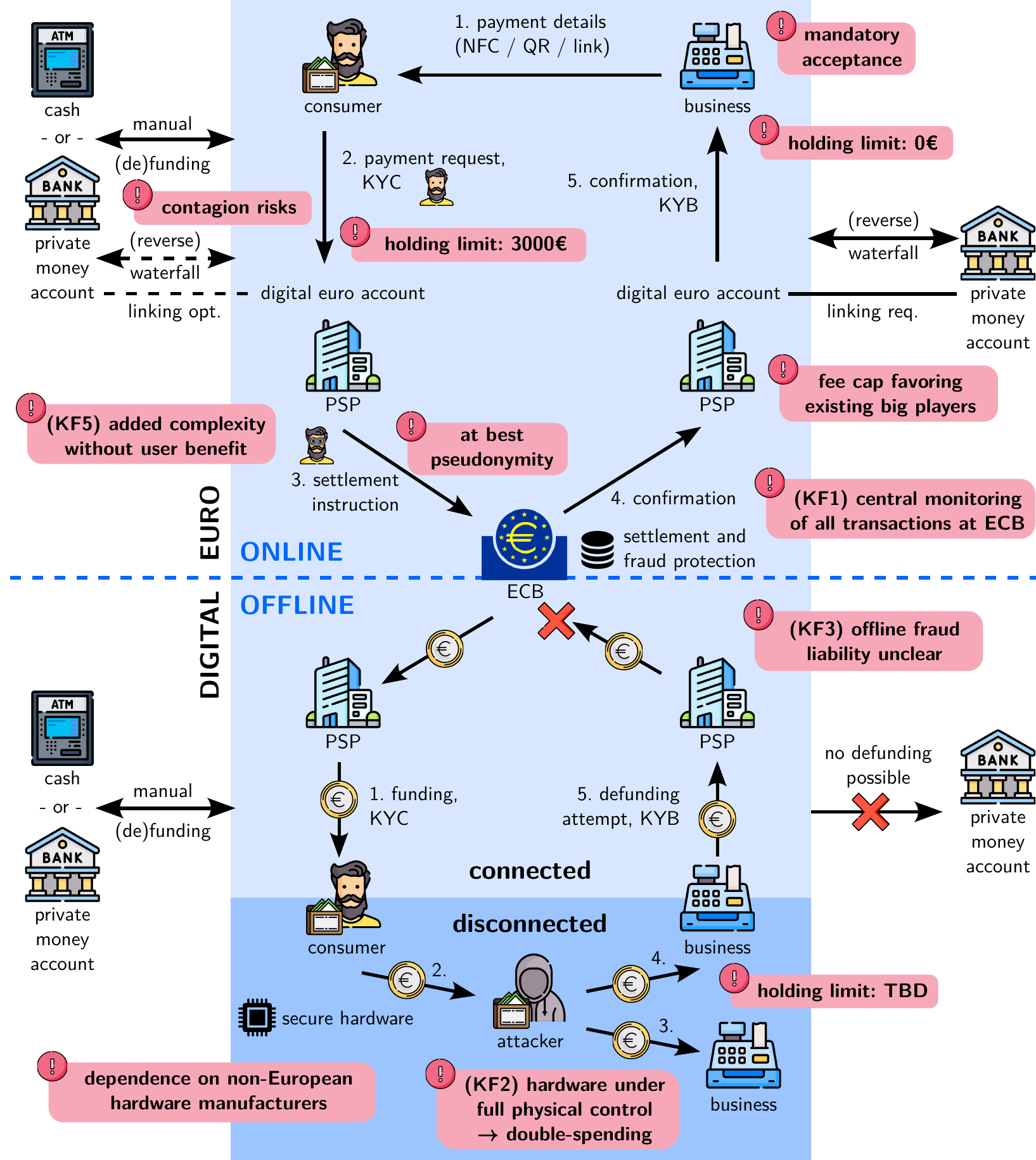}
	\caption{
		\textbf{The two versions of the digital euro} with highlighted design shortcomings, as described in the subsequent sections:
		The online version adds another layer of complexity to already existing payment systems,
		with no clear benefit for the user \KF{\Benefits},
		but with centralized control and monitoring of transactions at the ECB \KF{\Privacy}.
		The offline version is supposed to allow for transitive and completely anonymous payments
		and aims to prevent forgery by the use of supposedly ``secure hardware''.
		However, it is unlikely that the wallet hardware will withstand all possible attacks
		given that it is under complete physical control of the wallet owner as a potential attacker \KF{\Offline}.
		A successful attacker can double-spend an unlimited amount of times,
		as the validity of the payment bearer can only be checked as soon as the payee reconnects to the Internet.
		The offline version has no clear definition of liability in case of such fraud \KF{\Legal}.
	}
	\label{fig:problems}
\end{figure}

The ECB proposes ``the digital euro'', suggesting a single digital version of the euro.
In fact, its design encompasses two mostly separate parts as depicted in \autoref{fig:problems}:
The online version
	which adds another layer of complexity to the existing digital payment infrastructure;
and the offline version
	which is supposed to mostly resemble physical cash
	by providing transitive anonymity and operation with only occasional Internet connectivity.
Both follow a two-tiered architecture,
	where the ECB acts as the central trusted instance for issuance and settlement.
Participating payment service providers (PSPs) such as banks or public entities~\cite[Rec.~29]{ecDE2023}
	interact with end-users for identification and authentication
	adhering to legal know-your-customer (KYC) and know-your-business (KYB) requirements.
Using the digital euro is free of charge for end users,
	while PSPs are allowed (and expected) to charge fees up to a certain cap from merchants,
	who at the same time will be legally required to accept the digital euro~\cite[Rec.~16,20]{ecDE2023}.

\paragraph{Online version}
Users of the online version have to open
	a separate digital euro account (DEA) at a PSP.
Different to common bank accounts,
	these are interest free and subject to a holding limit
	to ensure financial stability of the contemporary commercial bank system~\cite[p.~8-9]{ecbProgPrepFirst2024}.
In order to ease user experience despite this holding limit,
	it is possible to link a commercial bank account to the DEA
	for automatic defunding in case the reception of a transaction would raise the balance above the holding limit (``waterfall''),
	and for automatic funding in case the amount to be paid exceeds the DEA balance (``reverse waterfall'')~\cite[p.~15,17]{ecbRulebookUpdate}.
The holding limit for non-business users is discussed to be around €3000~\cite{ecbHoldingLimit2023},
	while business users are not allowed to hold digital euros at all (with still unclear implications for offline usage),
	thereby making the linking of a commercial bank account a requirement for business users~\cite[p.~12-13]{ecbFinalInv2023}.
All online digital euro transactions are visible to the PSP,
	while the transaction data for the central settlement at the ECB is at best pseudonymized~\cite[p.~18]{ecbProgPrep2025}.
To ensure the validity of digital euro transactions,
	the ECB maintains a central database of all DEA balances~\KF{\Privacy}~\cite[p.~18-20]{ecbRulebookUpdate}.

\paragraph{Offline version}
The offline version~\cite[p.~4-6]{ecbProgPrepFirst2024} builds on a completely separate technical basis:
	a digital bearer (token) that represents a certain value,
	issued and validated at the ECB as a central instance.
Aiming to prevent digital copies of the bearer and double-spending, the bearer is
	stored on a dedicated device (e.g., smartphone or smartcard) supported by
	so-called ``secure hardware''---a separate piece of hardware for sensitive data
	promising protection against software and hardware attacks.
Those bearers are meant to be passed transitively between users without central settlement,
	and to not record any trace of the transactions,
	thereby providing full anonymity for both payer and payee.
However, Internet connectivity and identification is required
	for initial funding (conversion of cash or commercial bank money to digital euros)
	and final defunding (conversion of digital euros to cash or commercial bank money),
	among others to ensure the holding limit shared with the online version~\cite[p.~14,16]{ecbRulebookUpdate}.
Fraud such as double-spending can only be detected with a delay during defunding~\KF{\Offline},
	and the ECB leaves the question of liability in such cases open~\KF{\Legal}.

\section{Privacy: undercutting existing systems \KF{\Privacy}}
\label{sec:privacy}

The ECB compares the digital euro to cash and emphasizes privacy (\ref{q:privacy}),
	matching clear consumer preference for payment privacy in recent surveys~\cite{ecb2021survey,paymenthabits2022,paymenthabits2024}.
Unfortunately it fails to deliver:
	While the offline version indeed promises full transaction privacy with respect to the payment infrastructure,
	it remains unclear how an offline fraud incident~\KF{\Offline} would be handled without any recorded transaction history \KF{\Legal}.
Given the convenience functions of the online version such as automatic deposit and withdrawal,
	and given that Internet is available in many situations,
	people are likely to stick to the online version of the digital euro,
	probably in the illusion that their transaction data is private there, too.

However, the online digital euro mirrors the design of typical digital payment systems \KF{\Benefits}
	where all transactions are completely visible to the PSP,
	with at most organizational or legal, but no strong cryptographic safeguards against data misuse.
And, even worse, the ECB will maintain a central database of all online digital euro transactions,
	where it, as they state, ``would not be able to directly link (...) transactions to specific individuals'' (\ref{q:privacy}).
Published documents solely reference pseudonymization of individuals,
	i.e., using unique identifiers instead of real names and personal identities for digital euro accounts.
However, this still allows the creation of ``patterns of life'' and detailed insights into citizen's private lives,
	where it is enough to relate a single transaction with a certain individual to obtain their whole transaction history---%
	indeed not ``directly'', but with little effort~\cite{reidentifiability2015}.
This will enable an unprecedented level of easy mass surveillance
	and represent a high-value target for cyberattacks (\ref{q:cyberattacks})
	as payment data would no longer be siloed across thousands of organizations,
	databases, and incompatible formats~\cite{suerf2022aligny}.

The European Convention on Human Rights
	establishes privacy as a fundamental human right,
	which however is not absolute but subject to derogation~\cite[Art.~8]{coe1950humanrights}.
Any privacy-intrusive measure must pass a number of tests derived
	from the wording and subsequent interpretation of Article~8:
There needs to be a clear legal basis for the measure,
	remedies for breach of privacy must be established,
	and the measure must be both necessary and proportionate in a democratic society.
The privacy-intrusive nature of the online digital euro,
	where PSPs provision digital euro accounts to verified identities
	and can thus relate every single transaction to an individual,
	is commonly justified by the necessity of Anti-Money Laundering (AML) legislation.
This oversees the fact that alternative designs for digital payment systems exist,
	which provide KYC-compliant money inflow
	and income transparency through identifiable payees,
	to serve an equally good purpose in countering criminal activities,
	while offering anonymity on the payer's side~\cite{suerf2021moser,platypus};
	also referred to as ``asymmetric privacy''~\cite{tinn25}.
But even if one were to accept that the two tests of
	legal basis and necessity have been met by the draft regulation~\cite{ecDE2023},
	the digital euro would likely fail the test of proportionality.
The privacy risks implied by a central database
	which is already discussed being in reach of intelligence services and police forces~\cite{henning2024netzpolitik}
	are significantly disproportionate to the functionality or any other advantage gained for the citizen \KF{\Benefits}.
It is unfortunately not a given that all intelligence services of all EU member states will forever be
	trustworthy to not misuse their access to such a huge database for nefarious purposes.
Instead of ``privacy by design'', at this moment in time,
	the digital euro is promising ``less privacy through flawed design''.

\section{Security: ignoring obvious attack vectors \KF{\Offline}}

\begin{figure}[t]
	\def\svgwidth{1\textwidth}

\begingroup%
  \makeatletter%
  \providecommand\color[2][]{%
    \errmessage{(Inkscape) Color is used for the text in Inkscape, but the package 'color.sty' is not loaded}%
    \renewcommand\color[2][]{}%
  }%
  \providecommand\transparent[1]{%
    \errmessage{(Inkscape) Transparency is used (non-zero) for the text in Inkscape, but the package 'transparent.sty' is not loaded}%
    \renewcommand\transparent[1]{}%
  }%
  \providecommand\rotatebox[2]{#2}%
  \newcommand*\fsize{\dimexpr\f@size pt\relax}%
  \newcommand*\lineheight[1]{\fontsize{\fsize}{#1\fsize}\selectfont}%
  \ifx\svgwidth\undefined%
    \setlength{\unitlength}{346.44799805bp}%
    \ifx\svgscale\undefined%
      \relax%
    \else%
      \setlength{\unitlength}{\unitlength * \real{\svgscale}}%
    \fi%
  \else%
    \setlength{\unitlength}{\svgwidth}%
  \fi%
  \global\let\svgwidth\undefined%
  \global\let\svgscale\undefined%
  \makeatother%
  \begin{picture}(1,0.34501279)%
    \lineheight{1}%
    \setlength\tabcolsep{0pt}%
    \put(0,0){\includegraphics[width=\unitlength,page=1]{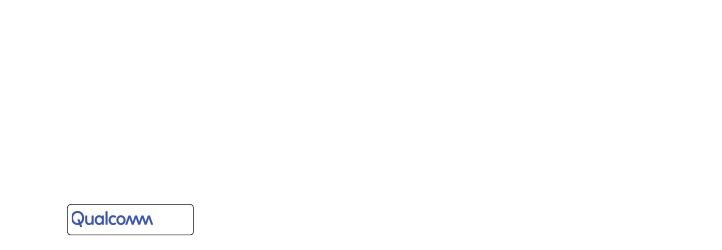}}%
    \put(0.21543115,0.03360253){\color[rgb]{0,0,0}\makebox(0,0)[lt]{\lineheight{1.25}\smash{\begin{tabular}[t]{l}\cite{arm2017boomerang}\end{tabular}}}}%
    \put(0,0){\includegraphics[width=\unitlength,page=2]{offline-timeline.pdf}}%
    \put(0.56613307,0.08856732){\color[rgb]{0,0,0}\makebox(0,0)[lt]{\lineheight{1.25}\smash{\begin{tabular}[t]{l}\cite{intel2020lvi,intel2020sgaxe}\end{tabular}}}}%
    \put(0,0){\includegraphics[width=\unitlength,page=3]{offline-timeline.pdf}}%
    \put(0.34965039,0.03372502){\color[rgb]{0,0,0}\makebox(0,0)[lt]{\lineheight{1.25}\smash{\begin{tabular}[t]{l}\cite{arm2017clkscrew}\end{tabular}}}}%
    \put(0,0){\includegraphics[width=\unitlength,page=4]{offline-timeline.pdf}}%
    \put(0.50395012,0.03357902){\color[rgb]{0,0,0}\makebox(0,0)[lt]{\lineheight{1.25}\smash{\begin{tabular}[t]{l}\cite{amd2019}\end{tabular}}}}%
    \put(0,0){\includegraphics[width=\unitlength,page=5]{offline-timeline.pdf}}%
    \put(0.42398865,0.08856726){\color[rgb]{0,0,0}\makebox(0,0)[lt]{\lineheight{1.25}\smash{\begin{tabular}[t]{l}\cite{samsung2017knox}\end{tabular}}}}%
    \put(0,0){\includegraphics[width=\unitlength,page=6]{offline-timeline.pdf}}%
    \put(0.16787832,0.08905342){\color[rgb]{0,0,0}\makebox(0,0)[lt]{\lineheight{1.25}\smash{\begin{tabular}[t]{l}\cite{arm2016alias,arm2016cache,zhang2016truspy}\end{tabular}}}}%
    \put(0,0){\includegraphics[width=\unitlength,page=7]{offline-timeline.pdf}}%
    \put(0.35532942,0.24699274){\color[rgb]{0,0,0}\makebox(0,0)[lt]{\lineheight{1.25}\smash{\begin{tabular}[t]{l}\cite{sim2019}\end{tabular}}}}%
    \put(0,0){\includegraphics[width=\unitlength,page=8]{offline-timeline.pdf}}%
    \put(0.51060396,0.24720308){\color[rgb]{0,0,0}\makebox(0,0)[lt]{\lineheight{1.25}\smash{\begin{tabular}[t]{l}\cite{smartcard2020}\end{tabular}}}}%
    \put(0,0){\includegraphics[width=\unitlength,page=9]{offline-timeline.pdf}}%
    \put(0.40593556,0.30216357){\color[rgb]{0,0,0}\makebox(0,0)[lt]{\lineheight{1.25}\smash{\begin{tabular}[t]{l}\cite{nuvotoninfineon2018}\end{tabular}}}}%
    \put(0,0){\includegraphics[width=\unitlength,page=10]{offline-timeline.pdf}}%
    \put(0.92984328,0.03362909){\color[rgb]{0,0,0}\makebox(0,0)[lt]{\lineheight{1.25}\smash{\begin{tabular}[t]{l}\cite{intel2024}\end{tabular}}}}%
    \put(0,0){\includegraphics[width=\unitlength,page=11]{offline-timeline.pdf}}%
    \put(0.90235344,0.08856729){\color[rgb]{0,0,0}\makebox(0,0)[lt]{\lineheight{1.25}\smash{\begin{tabular}[t]{l}\cite{amd2025, amd2025b}\end{tabular}}}}%
    \put(0,0){\includegraphics[width=\unitlength,page=12]{offline-timeline.pdf}}%
    \put(0.86054978,0.30211543){\color[rgb]{0,0,0}\makebox(0,0)[lt]{\lineheight{1.25}\smash{\begin{tabular}[t]{l}\cite{infineon2024}\end{tabular}}}}%
    \put(0,0){\includegraphics[width=\unitlength,page=13]{offline-timeline.pdf}}%
    \put(0.88691555,0.2465412){\color[rgb]{0,0,0}\makebox(0,0)[lt]{\lineheight{1.25}\smash{\begin{tabular}[t]{l}\cite{tpm2023}\end{tabular}}}}%
    \put(0,0){\includegraphics[width=\unitlength,page=14]{offline-timeline.pdf}}%
    \put(0.23625647,0.24727314){\color[rgb]{0,0,0}\makebox(0,0)[lt]{\lineheight{1.25}\smash{\begin{tabular}[t]{l}\cite{infineon2017}\end{tabular}}}}%
    \put(0,0){\includegraphics[width=\unitlength,page=15]{offline-timeline.pdf}}%
    \put(0.64405814,0.0337338){\color[rgb]{0,0,0}\makebox(0,0)[lt]{\lineheight{1.25}\smash{\begin{tabular}[t]{l}\cite{intel2023sgx}\end{tabular}}}}%
    \put(0,0){\includegraphics[width=\unitlength,page=16]{offline-timeline.pdf}}%
    \put(0.79127542,0.0333083){\color[rgb]{0,0,0}\makebox(0,0)[lt]{\lineheight{1.25}\smash{\begin{tabular}[t]{l}\cite{amd2023}\end{tabular}}}}%
    \put(0,0){\includegraphics[width=\unitlength,page=17]{offline-timeline.pdf}}%
    \put(0.75033382,0.08805098){\color[rgb]{0,0,0}\makebox(0,0)[lt]{\lineheight{1.25}\smash{\begin{tabular}[t]{l}\cite{arm2023}\end{tabular}}}}%
    \put(0,0){\includegraphics[width=\unitlength,page=18]{offline-timeline.pdf}}%
    \put(0.54737115,0.30216356){\color[rgb]{0,0,0}\makebox(0,0)[lt]{\lineheight{1.25}\smash{\begin{tabular}[t]{l}\cite{st2020}\end{tabular}}}}%
    \put(0,0){\includegraphics[width=\unitlength,page=19]{offline-timeline.pdf}}%
    \put(0.71036969,0.30250353){\color[rgb]{0,0,0}\makebox(0,0)[lt]{\lineheight{1.25}\smash{\begin{tabular}[t]{l}\cite{nxp2021}\end{tabular}}}}%
    \put(0,0){\includegraphics[width=\unitlength,page=20]{offline-timeline.pdf}}%
    \put(0.70194435,0.24717288){\color[rgb]{0,0,0}\makebox(0,0)[lt]{\lineheight{1.25}\smash{\begin{tabular}[t]{l}\cite{atecc2022}\end{tabular}}}}%
    \put(0,0){\includegraphics[width=\unitlength,page=21]{offline-timeline.pdf}}%
    \put(0.14541157,0.12873844){\color[rgb]{0,0,0}\makebox(0,0)[lt]{\lineheight{1.25}\smash{\begin{tabular}[t]{l}2015\end{tabular}}}}%
    \put(0.52064824,0.12873844){\color[rgb]{0,0,0}\makebox(0,0)[lt]{\lineheight{1.25}\smash{\begin{tabular}[t]{l}2020\end{tabular}}}}%
    \put(0.89588494,0.12873844){\color[rgb]{0,0,0}\makebox(0,0)[lt]{\lineheight{1.25}\smash{\begin{tabular}[t]{l}2025\end{tabular}}}}%
    \put(0,0){\includegraphics[width=\unitlength,page=22]{offline-timeline.pdf}}%
    \put(0.05158787,0.07038109){\color[rgb]{0,0,0}\rotatebox{90}{\makebox(0,0)[t]{\lineheight{1.25}\smash{\begin{tabular}[t]{c}TEE\end{tabular}}}}}%
    \put(0.0366477,0.26157866){\color[rgb]{0,0,0}\rotatebox{90}{\makebox(0,0)[t]{\lineheight{1.25}\smash{\begin{tabular}[t]{c}SE\\(TPM, eSIM)\end{tabular}}}}}%
    \put(0,0){\includegraphics[width=\unitlength,page=23]{offline-timeline.pdf}}%
    \put(0.116244,0.32583068){\color[rgb]{0,0,0}\makebox(0,0)[lt]{\lineheight{1.25}\smash{\begin{tabular}[t]{l}physical attack\end{tabular}}}}%
    \put(0.11732934,0.29144745){\color[rgb]{0,0,0}\makebox(0,0)[lt]{\lineheight{1.25}\smash{\begin{tabular}[t]{l}software attack\end{tabular}}}}%
  \end{picture}%
\endgroup%

	\vspace{-0.6cm}
	\caption{{\bf Secure offline payments require securing a piece of hardware against its owner.}
		History suggests this is impractical for consumer-grade hardware \KF{\Offline}.
		Each box in the figure represents an attack that broke the security promises made by its vendor.
		Compromising Trusted Execution Environments (TEEs) has proven possible with software-only attacks,
		while successful attacks on Secure Elements (SEs) mostly required physical access to the device.
		Mitigations thus often require hardware changes.
		We cannot demand people to purchase new devices anytime a ``secure hardware'' proves vulnerable to attacks.}
	\label{fig:offline}
	\vspace{-0.3cm}
\end{figure}

The ECB promises ``secure instant payments (...) even when you have (...) no network reception'' (\ref{q:offline}),
	i.e., reliable offline operation without double-spending risks.
This contradicts the mathematically proven CAP theorem~\cite{cap},
	which states that no distributed system can be
	partition-tolerant (being disconnected), available (still work), and consistent (without double-spending) at the same time.
The ECB hopes to overcome mathematics with ``secure hardware'',
	in this case meant to ``protect the information stored on the device'' from its owner~\cite{ecbDEoffline2024},
	ignoring the fact that hardware security history has made it evident that
	consumer-grade hardware eventually cannot withstand physical attacks (cf.~\autoref{fig:offline}).
Focusing on dedicated devices like secure elements (SEs) and eSIMs
	and excluding Trusted Execution Environments (TEEs), as recently declared by the ECB in \cite[p.~16]{ecbProgPrep2025},
	does not solve this critical issue:
While physical attacks are typically needed to compromise such hardware,
	the potential return by unlimited digital euro duplication
	will likely outweigh the necessary investment costs.
Note that there is a key difference in the attack scenario on hardware for the offline digital euro
compared to other typical usages of ``secure hardware'' such as payment card details or private keys for two-factor authentication:
Owners of such hardware have no interest in regaining information---as they already own it---%
	and being resistent to remote software or unnoticeably fast hardware attacks is sufficient.
In contrast, the offline digital euro requires protecting information from the physical owner of the device
	who has unlimited time to break it.

The ECB's approach of just hiding details about the used proprietary technology~\cite{markpersonal} is promising ``security by obscurity'',
	which contradicts Kerckhoffs' principle---a fundamental concept in cryptography which asserts that
	the security of a system should not depend on the secrecy of its algorithm~\cite{kerckhoffs1883}.
G+D's platform, the selected solution for the offline digital euro implementation~\cite{gdOffline2024}
	which proposes exactly such a system,
	has been ranked as ``low'' in a recent survey due to secrecy being the main line of defense against double-spending~\cite{chavanette2024}.
By proposing a ``forgery check during defunding''~\cite{ecbDEoffline2024},
	the ECB acknowledges the risk of hardware-level attacks or implementation bugs,
	but thereby contradicts the claim of offline payments being ``safe and instant'',
	and further puts the promised complete anonymity of offline transactions into question.

The ``reverse waterfall'' mechanism, the automatic digital euro funding without user intervention,
	poses another significant security threat as a compromised digital euro account
	could be used to obtain unlimited money from the linked commercial bank account.
The ECB promises ``state-of-the-art technologies'' to counter cyberattacks on the digital euro (\ref{q:cyberattacks}).
This term is doubly misleading:
	First, no other large-scale deployment of digital offline payments exists today---as the ECB asserts in~\cite[p.~16]{ecbProgPrep2025}.
	Second, it does not imply the application of the most modern or secure solutions available,
	but rather just ``what everybody else does''.
The central database of all (online) digital euro transaction being such a high-value target for cyberattacks \KF{\Privacy},
	this is most probably not enough.

\section{Legal and financial: vague liability and unarticulated risks \KF{\Legal}}

The ECB has not clarified compensation mechanisms for fraud victims in offline scenarios.
It barely states that
	``either the PSP, the merchant or, in some cases, the consumer would be liable'',
	never the ECB itself~\cite{ecbFinalInv2023}.
It also remains unclear how the ECB will react
	when the double-spending protection has been overcome,
	enabling unlimited digital euro creation
	for every owner of wallets relying on the affected hardware \KF{\Offline}.
It is worth highlighting that the existing threat of counterfeited cash is limited
	by physical challenges of production and distribution for the counterfeiter;
in the digital world, though, limiting the ability to copy data,
	once it becomes accessible, is impossible.
Furthermore, the size of potential financial damage
	when protections have been overcome,
	scales linearly with the amount of transactions within the system,
	not the (smaller) aggregate amount stored in it.
As this eventually poses a financial threat to the (digital) euro (\ref{q:financial-stability}),
	the offline functionality would need to be
	disabled for the given type of payer's devices---%
	which is only possible when all payee's devices become connected---%
	until the affected hardware is replaced.
Apart from not mentioning the security and financial risks connected to the offline version,
	the ECB also keeps quiet about the additional costs and complex logistics
	incurred by such hardware breaches.

Framing the digital euro as ``stable and reliable'' (cf.~\ref{q:crypto}) is an oversimplification.
	Being legal tender and backed by a central bank may make it a counter-party risk-free store of value for citizens,
	but the underlying digital payment system is still exposed to a broad range of risks,
	as discussed in this article.
The ECB, as the responsible public operator of the digital euro,
	and ultimately the general public, must take these risks into consideration.

\section{Economical: unclear incentives and market distortions \KF{\Economic}}

While the ECB emphasizes the advantages of the digital euro for merchants (\ref{q:merchants}),
	it fails to mention the likely substantial costs of integration into their business systems.
The legally enforced roll-out throughout virtually all points of sale in the euro area
	will create a high demand for qualified IT integrators,
	allowing them to charge a high premium over normal integration costs.

Supervised intermediaries---the PSPs---are expected to
	``perform all end-user services'' (\ref{q:psps}).
This entails operation and on-boarding including costly KYC and KYB requirements,
	which need to be reimbursed solely by the fees charged to merchants (\ref{q:psps-compensation}),
	since the digital euro is promised to be ``free for basic use'' for consumers.
The proclaimed ``economic incentives comparable to other digital means of payments'' for PSPs (\ref{q:psps})
	beg the question of where benefits for the PSPs will come from.
To cope with costs, one (generally undesirable) option might be to compromise security
	if fraud liability stayed at the ECB---which is unclear \KF{\Legal}.
Another option for commercial banks	could be to earn interest
	on user-facing digital euros under their management via the ECB's deposit facility,
	by offering an extra euro account tightly integrated with the digital euro where users could earn interest.
Since non-bank PSPs have no access to the deposit facility, this would be an unfair advantage for commercial banks~\cite{guetschow2025deregulation}.

While the ECB hopes for the digital euro to make ``the European payments landscape more competitive and
innovative'' (\ref{q:autonomy}),
	it remains unclear how new or smaller PSPs will be able to compete with big players having a pre-existing user base.
The uniformity of the service offering and the cap on allowed fees favor market oligopolization among digital euro PSPs.
At the same time, a legally mandated digital euro could sideline and displace established national digital payment systems,
	leaving less consumer choice and eventually stifling innovation.

\section{Utility and costs: unclear benefits for society \KF{\Benefits}}

\begin{table}
	\renewcommand{\arraystretch}{1.5}
	\newlength{\fw}
	\setlength{\fw}{4cm}
	\centering
	\sf
	\footnotesize
	\begin{tabular}{l c c c c c c c c}
	\toprule
	& \multicolumn{5}{c}{Digital Payment System} & \multicolumn{2}{c}{Digital Euro}
	\\
	\cmidrule(lr){2-6}
	\cmidrule(lr){7-8}
	Feature          & Credit Card  & PayPal  & Wero    & Bitcoin & GNU Taler   & Online  & Offline & Cash    \\ \midrule
	Online operation & \y      & \y      & \y      & \y      & \y      & \y      & \n      & \n      \\
	Offline operation & (\y)    & \n      & \n      & \n      & \n      & \n      & \y      & \y      \\ \dhline
	Payer anonymity	 & \n      & \n      & \n      & (\n)    & \y      & \n      & \y      & \y      \\
	Payee anonymity  & \n      & \n      & \n      & (\n)    & \n      & \n      & \y      & \y      \\ \dhline
	AML compliance   & \y      & \y      & \y      & (\n)    & \y      & \y      & (\n)      & (\n)      \\ \dhline
	Security\fn{1}     & \n  & \n  & \n & \n  & \y  & \n  & \n  & \y  \\
	\ no illicit remote access & \n  & \n  & \n  & \n  & \y  & \n  & \y  & \y  \\
	\ no double-spending  & \y  & \y  & \y  & \y  & \y  & \y  & \n  & \y  \\
	\ reasonable finality & \y  & \y  & \y  & \n  & \y  & \y  & \n\fn{2}  & \y  \\ \dhline
	Prior funding needed  & \n      & \n      & \n      & \y      & \y      & \n\fn{3}  & \y      & \y      \\ \dhline
	Libre software   & \n      & \n      & \n      & \y      & \y      & \n      & \n      & -       \\
	\bottomrule
	\end{tabular}

	\hskip 1em \vskip 1ex \raggedright
	\hskip 1em \fn{1}security requires all three aspects
	\hskip 1em \fn{2}forgery check at deferred online defunding
	\hskip 1em \fn{3}using waterfall approach

	\vskip 1em

	\caption{{\bf Comparison of the two digital euro versions to other digital payment systems and cash.}
		The online version differs significantly from the offline one in terms of both privacy and security.
		Since the former is designed with an architecture similar to already existing third-party payment systems \KF{\Benefits},
		it matches those quite closely with respect to the available features,
		but fails to deliver on additional possible benefits such as payer anonymity.
		The offline version seems to resemble cash most, but contains inherent security flaws \KF{\Offline}
		and requires Internet connectivity for a prior manual funding operation.
		}
	\label{tab:comparison}

	\vspace{-1cm}

\end{table}

The ECB claims various advantages of the digital euro,
	including no costs for end-users,
	universal acceptance in physical shops and online,
	support for person-to-person payments,
	``the highest level'' of security and privacy,
	resilience to cyberattacks and technical disruptions,
	technological independence contributing to Europe's strategic autonomy,
	and non-reliance on Internet connectivity (\ref{q:resilience}, \ref{q:autonomy}, \ref{q:benefits}).
The only tangible advantage compared to cash is the ability to pay digitally and online,
	while the comparison of the digital euro design to existing digital payment systems in \autoref{tab:comparison} only leaves
	the legally enforced acceptance throughout the euro area as a clear improvement for the user.
Many payment systems can be used for person-to-person payments, online, and at points of sales;
	and are also free of charge for consumers, with merchants covering the fees.
Being a central bank liability makes no effective difference to the vast majority of Europeans,
	given that bank deposits are generally insured up to €100,000,
	while digital euro holdings are expected to be limited to at most €3,000~\cite[p.~25]{ecbProgPrep2025}
	as well as interest-free and thereby less attractive for savings (\ref{q:financial-stability}).
The claims of improved resilience to cyberattacks lack supporting evidence
	in terms of concrete measures provided in public documents~(\ref{q:cyberattacks}).
Using the offline version in case of technical disruptions such as power outages
	is only possible after a manual funding process which still needs Internet connectivity.
Given the convenience option of automatic funding in the online version,
	it seems questionable to which extent people would have offline digital euros available when their Internet goes down.
While the digital euro could reduce the dependence on non-European payment providers,
	the offline version still depends on proprietary, predominantly non-European hardware.
Truely embracing open standards and free software~\cite{floss} instead of relying on proprietary technology
	would allow for easier integration with alternative European hardware and end-user devices
	and help to establish trust in the system \KF{\Process}.
The high risks of the offline version have been articulated for \KF{\Offline},
	and the online version offers even less privacy than other digital payment systems
	due to the existence of a central transaction database \KF{\Privacy}.

While having little clear advantages,
	the digital euro project brings significant disadvantages for euro area citizens:
First, with an initial project budget of €1.3 billion (cf.~\ref{q:costs-system}),
	it is far from being a ``cheap (...) form of public money'' (\ref{q:legislation}),
	especially when compared to more innovative and cost-efficient digital payment systems~\cite{tsys},
	with developments costs of less than 1\% of that amount (\ref{q:costs-system}).
The high additional costs for mandatory roll-out to all merchants, support, and onboarding \KF{\Economic} are not even included in that number.
In fact, all costs of the digital euro are eventually borne by the citizens of the euro area,
	whether they want to use it or not,
	contrary to the ECB's claim that the digital euro would be ``free for basic use by individual users'' (\ref{q:costs-users}).
Second, while the ECB and the European legislation may try to avoid it (\ref{q:cash}),
	bringing yet another digital payment system to the market will likely further erode the use of physical cash,
	which is the most power-outage-resilient, inclusive, accessible, and privacy-preserving payment system we know of.

\section{Governance and transparency: exclusionary process
    and missed opportunities of an open distribution model \KF{\Process}}
\label{sec:process}

The proclaimed openness of the digital euro design finding process (\ref{q:consultation}) is questionable,
	given that many of the core design features---%
	such as the online-offline separation, the waterfall concept, and the reliance on ``secure hardware''---%
	have been defined before public consultations began~\cite{digitaleuro2020}.
On the contrary, the ECB has largely ignored expert advice
	on how to design for privacy, security, and usability~\cite{suerf2022aligny,suerf2021moser,platypus,uhlig2023privacy}.
The ECB engagement with private companies (\ref{q:consultation}) has excluded small and medium enterprises
	by setting high thresholds for potential participants,
	requiring, among others, a yearly average total net turnover of €100,000,000~\cite{ecbTender0078480,ecbTender009488}.
The digital euro rulebook (\ref{q:rulebook}) is developed behind closed doors
	and leaves no room for questioning fundamental digital euro design choices impacting privacy~\KF{\Privacy}, security~\KF{\Offline}, and usability~\KF{\Benefits}.


\begin{table}[t]
\centering
\begin{tabular}{l c c c c}
\toprule
& \multicolumn{4}{c}{Distribution model} \\
\cmidrule(lr){2-5}
User Right           & Proprietary & Source-available & Open-core & Free software \\ \midrule
Execute              & \y    & \y      & \y    & \y    \\ \dhline
Analyze/integrate    & \n    & (\y)    & \y    & \y    \\ \dhline
Modify/improve       & \n    & \n      & (\y)  & \y    \\ \dhline
Fully self-determine$^\dagger$ & \n    & \n      & \n    & \y    \\
\bottomrule
\end{tabular}
\caption{{\bf Overview of the different distribution models for software
	and the rights that they provide to the user.}
In the context of CBDC's, ``user'' refers to (a) the end-user,
	who will pay in digital euros by using software,
	(b) the merchants, who will need software systems to accept digital euros,
	and also (c) to the ECB itself and associated third-parties,
	who will run the backbone of the digital euro system with software.\\
 {\footnotesize $\dagger$ Informational self-determination requires control over the software you use.}}
\label{tab:sw-models}
\vspace{-1cm}
\end{table}

While the ECB emphasizes the need for open standards and public infrastructure
	to foster inter-operation of systems across Europe (\ref{q:autonomy}) and beyond,
	it misses on the opportunity to specify the distribution model for the software
	that will be run by the ECB and third-party providers (for the backbone of the digital euro),
	the merchants (in order to accept payments in digital euro)
	and the end-users (in order to pay).
The distribution models of the software/hardware co-design of any digital solution
	can be broadly categorized into four types.
Ordered from most restrictive to most freedom respecting, these are (see also \autoref{tab:sw-models}):
\begin{enumerate}
	\item \emph{Proprietary}, where the design and implementation is kept as a secret,
		and users or other third parties are legally forbidden to inspect the internals;
	\item \emph{Source-available}, where part of the design and/or implementation is available
		to allow for high-level analysis and restricted integration with third-party services,
		potentially only to certain actors;
        \item \emph{Open-core}, building on a
                system where the base implementation is publicly available for anyone to view, modify and redistribute, but
                where key features are kept proprietary; and
        \item \emph{Fully free software},
                which emphasizes the user's human right to have full control over the software run on their devices
                and thus excludes proprietary extensions or other forms of digital shackles such as hardware locks.
		In particular, ``free'' in this context refers to ``freedom'', and not to the price~\cite{floss}.
\end{enumerate}

\vspace{-.3em}

The option for third-parties to analyze, re-use and improve a certain design and implementation
	allows for broader acceptance (e.g., in other parts of the world) and joint innovation~\cite{raymond1999cathedral}.
It also stimulates competition and improves user choice
	by enabling the provision of end-user services and devices by a larger number of providers.
However, the design of the offline digital euro is expected
	to rely on its proprietary nature for security~\cite{markpersonal}---contradicting Kerckhoffs' principle~\KF{\Offline}.
The ECB has not communicated any clear licensing model for the online version yet,
	but reserved all the rights on the implementation for themselves in the public tenders~\cite{ecbTender009488}.

\section{Recommendations}
\label{sec:recommendations}

The previous sections, and in more detail \autoref{sec:appendix:faq}, question many of the design decisions
	of the digital euro in response to the official FAQ by the ECB.
To summarize, we give a short list of recommendations
	for a successful digital version of the euro or any other CBDC:

\begin{enumerate}
	\item \textbf{Acknowledge the impracticability of the offline version.}
		The ECB should focus on the online version,
		make it a single, privacy-preserving digital euro system~\KF{\Privacy},
		instead of spending money and time on a design idea
		based on the unrealistic assumption of unbreakable hardware~\KF{\Offline}.
		The ECB should acknowledge the fact that a digital currency
		will never be as resilient as physical cash to extraordinary events like blackouts.
		Also, without the offline version, related questions of liability can safely be ignored~\KF{\Legal}.
	\item \textbf{Provide a real advantage to users.}
		The digital euro should provide a compelling, unique selling point to be successful~\KF{\Benefits},
		instead of merely adding another layer to the existing payment infrastructure
		under centralized ECB oversight~\KF{\Privacy}.
		One such advantage could be cash-like payer anonymity for online transactions.
	\item \textbf{Convince instead of enforce.}
		If the digital euro provided a real advantage,
		it would not be necessary to put pressure on merchants to accept it with the associated additional costs~\KF{\Economic}.
		Thus, applicable legislative plans should be changed
		to make the acceptance of digital euros completely voluntary.
	\item \textbf{Build on existing solutions and Free/Libre Software.}
		Given that explicit design goals of the digital euro include
		user trust in the payment system,
		interoperability between participating payment service providers,
		and being a role model for CBDCs worldwide,
		the digital euro should be based on open designs
		and Free/Libre Open Source Software (FLOSS)
        instead of proprietary technology~\KF{\Process} and security by obscurity~\KF{\Offline}.
\end{enumerate}

Overall, unless clear and uncontroversial benefits of the digital euro
are established, we advise against the introduction of the digital euro as currently proposed.

\section*{Acknowledgments}

\autoref{sec:appendix:faq} takes inspiration from the well-known C++ FQA~\cite{cppfqa}
	that takes an outsider's perspective on the rosy FAQ
	on the C++ programming language written by the C++ core community.
We thank Emmanuel Benoist and Christian Grothoff for
	extensive inspiration, editing and support.
We thank Jens Palsberg, Richard Stallman and Tanja Lange
	for constructive feedback.
This work was supported by the German Federal Ministry of Education
    under grant \href{https://concretecontracts.codeblau.de}{ConcreteContracts}.
The opinions, findings, and conclusions
    or recommendations expressed are those of the authors
    and do not necessarily reflect those of any of the funding agencies.

\bibliographystyle{ACM-Reference-Format}
\bibliography{biblio-defqa}

\appendix

{
\section{Questioning the Digital Euro FAQ}
\label{sec:appendix:faq}
\renewcommand{\thesubsection}{Q\arabic{subsection}}

The ECB provides answers to 27 questions on the digital euro
	in its official FAQ, last updated on March 5th, 2026~\cite{defaq2025}.
We question the answers
	based on our research and analysis of the publicly available information about the digital euro.
Some answers are oversimplifying the reality,
	some pass over unpleasant facts,
	and most of them reveal design problems with the digital euro.

For each question, we first quote the ECB's answer before then discussing it.

\subsection{Why does Europe need the digital euro?}
\label{q:resilience}

The ECB answers:
\begin{quote}\scriptsize\sf

In a world where digital payments are rapidly becoming the norm, the use of
cash is declining and the shift towards online shopping is accelerating.
The digital euro would be a digital form of cash, giving consumers access to
central bank money in digital form, complementing banknotes and coins.

The digital euro would make people’s lives easier by providing
something that does not currently exist: a digital means of payment
universally accepted throughout the euro area, for payments in
shops, online or from person to person. Like cash, the digital euro
would be accessible, free to use when making or receiving payments,
and have legal tender status.

Moreover, the digital euro would preserve the monetary sovereignty of
the euro area by boosting the efficiency of the European payments
ecosystem as a whole, fostering innovation and increasing its resilience
to cyberattacks and technical disruptions.

\end{quote}

The ECB observes the need of cash-like means of payment for the digital realm,
	and claims cash-like character for the digital euro.
But in its comparison of the digital euro to cash,
	the ECB ignores essential properties of cash,
	such as anonymity (cf.~\ref{q:privacy}) and independence of complex technologies (cf.~\ref{q:userbase}).
On the contrary, the digital euro is far from being as inclusive as cash (cf.~\ref{q:inclusive}).

The notion that the digital euro would be ``free to use'' oversimplifies the reality.
As discussed further in \ref{q:costs-users},
the costs associated with the implementation and maintenance of a digital currency are significant \KF{\Economic}
and in the end will be covered by the citizens of the euro area.
The legal tender status paired with a holding limit of €3,000 (cf.~\ref{q:financial-stability})
offers no tangible benefits over commercial bank money with deposit insurance up do €100,000 (cf.~\ref{q:benefits}).

The claim that the digital euro would
	increase ``resilience to cyberattacks or technical disruptions''
	unfortunately lacks support in published documents.
Indeed, it may not be surprising if a central database of all digital euro transactions
	became a prime target for cyberattacks (cf.~\ref{q:cyberattacks}).
Moreover, a digital currency system which depends on electronic infrastructure
	would in any case face significant challenges during technical disruptions such as power or Internet outages,
	particularly since converting online digital euros to offline versions
	requires Internet connectivity \KF{\Offline}.
Unlike physical cash, which remains usable in such scenarios,
	a fully digital system may fail if the infrastructure goes down.

\subsection{How could the digital euro boost Europe’s strategic autonomy?}
\label{q:autonomy}

\begin{quote}\scriptsize\sf

The digital euro would offer a pan-European payment solution,
available throughout the euro area, under European governance
and operated by European providers.

Digital payments in the euro area remain fragmented, differing
by country and by use case. More than half of all national markets
in the euro area do not have national digital payment solutions for
payments in shops, and those that do exist mainly cater only to national
markets and specific use cases. This means consumers have to rely on a
small number of non-European companies that dominate the market.
The digital euro would help reduce Europe’s dependence on non-European
payment service providers, and offer people the choice of using European solutions.

Through open standards and by providing a public infrastructure for digital payments,
the digital euro could enable providers to easily scale up to pan-European solutions,
thereby making the European payments landscape more competitive and innovative.

Overall, the digital euro could turn Europe into a global front-runner in digital finance,
where innovation serves the public good.

\end{quote}

While the digital euro could indeed by law reduce the dependence
	on non-European payment systems and technologies,
	it is worth noting that	its offline functionality relies on proprietary hard- and software
	from mostly non-European phone and hardware manufacturers.
Such proprietary design is also in contrast to the mention of ``open standards'' added in a recent revision of the FAQ.
Since neither the ECB nor the draft regulation establish a meaning for this term,
	it remains unclear what it effectively entails.

The European payment market is already dominated by two major companies, WorldLine and G+D,
	both of which are also key commercial players for the digital euro~\cite{worldlineDE2023,gdOffline2024,ecb2025tenderawards}.
It is thus questionable to which extent the digital euro
	would foster competition instead of consolidating the market dominance of few global stakeholders.
As a precedence, the thresholds for contribution to the digital euro design have been set very high by the ECB,
	effectively narrowing the potential group of contributors to very few big companies
	(cf.~\ref{q:consultation}).

On the contrary, the mere existence of the digital euro as government project
	with presumably low fees is expected to discourage investment in
	payment technologies by private actors~\cite[p.~12]{effects2023}.
Competing with a government-subsidized service is rarely viable in the commercial sector.
As a result, the digital euro project may actually stifle competition and innovation,
	and harm private European initiatives already existing today that aim at pan-European coverage (e.g., Wero)~\cite{navarette2025position}.
On the other hand, the published documents on the digital euro~\cite{ecDE2023,ecbFinalInv2023}
	do not contain evidence of innovative technology to improve efficiency,
	resilience or fraud protection.

The digital euro is also unlikely to be a ``global front-runner'' in the realm of CBDCs.
This is not only because it lags behind other global initiatives
	(such as the Chinese digital yuan, the Indian digital rupee, the Japanese digital yen,
	or the Brazilian digital currency Drex and instant payment system Pix)
	but also because it is designed to be based on proprietary European technology.
This creates a barrier for adoption in other countries,
	as they must weigh the risk of potential future sanctions.
Consequently, most large countries today favor either domestic or open technologies~\cite{chavanette2024}.
To become a global player, a CBDC technology would need to be fully free/libre and open-source software \KF{\Process},
	which is incompatible with the proprietary nature of the digital euro.

\subsection{Why would people want to use the digital euro?}
\label{q:benefits}

\begin{quote}\scriptsize\sf

The digital euro would be a payment solution for every occasion,
for use anytime and anywhere in the euro area  – just like cash,
but ready for the digital age. It would be a
universally accepted digital means of payment that consumers could
use free of charge in shops, online or from person to person. It
would give people the option to pay digitally, while still using a
public means of payment. And it would be available both online and
offline.

The digital euro would be designed to provide the highest level of
security and privacy, in compliance with the rigorous standards of
the European Union, which has the strongest security and privacy
laws in the world.

The Eurosystem would not identify people based on their payments.
Moreover, personal transaction details from offline
digital euro payments would be known only to the payer and the payee.

The digital euro would be safe and, as a public good, guarantee access
for all consumers, no matter where they live or their level of digital
or financial skills. The digital euro would accommodate the needs of people with
disabilities and those with no access to a bank account,
ensuring that no one is left behind.

To ensure the digital euro would be usable and accessible throughout
the euro area, the proposed digital euro Regulation presented by the
European Commission foresees mandatory acceptance by any merchant who
accepts digital payments, and mandatory distribution by banks to their
customers.

\end{quote}

In its current design, it is unclear what benefits the digital euro
	would bring to people in practice \KF{\Benefits}.
Contemporary cash and debit card payments already
	are generally free of charge for consumers---%
	with merchants covering the fees---and
	debit cards can work online and offline~\cite{offlinedebit2020kagan},
	in the latter case posing a financial risk for merchants
	as the transaction may fail at a later time.
Given that bank deposits are generally insured up to €100,000
	there is no effective difference between debit cards and
	central bank money for the vast majority of Europeans,
	as long as we assume the deposit insurance can withstand a crisis.
And for rich Europeans with savings larger than €100,000,
	the €3,000 cap on digital euro holdings~\cite{ecbHoldingLimit2023}
	has a negligible effect of 3\% in added securities,
	compared to simply opening another retail bank account.

The ECB answer claims that the digital euro
	will ``provide the highest level of security and privacy'',
	but the current design proves this statement wrong:
Fully anonymous transactions are only part of the offline version,
	which will likely not be able to hold up to this promise \KF{\Offline},
	as further detailed in \ref{q:offline}.
On the other hand, the online version of the digital euro will offer even less consumer privacy
	compared to contemporary payment methods \KF{\Privacy},
	as further discussed in \ref{q:privacy}.
It is very telling that the ECB, in a recent revision,
	slightly changed the wording from ``The Eurosystem would not be able to identify people based on their payments.''
	by removing ``be able to''~\cite{defaq2025old}.

The security of the offline version relies on secret proprietary technology \KF{\Offline},
	against well-established consensus among cryptographers (cf.~\ref{q:offline}).
This approach leaves the claim of a ``safe'' digital euro questionable
	and undermines the potential benefits of a ``public'' service,
	as users would remain dependent on proprietary digital payment technology,
	limiting transparency and fostering reliance on closed systems.

The current plans for the digital euro
	largely involve users opening a digital euro account
	and additionally linking it to a commercial bank account for the ``waterfall''~\cite{ecbProgInv2023}.
Thus, using the digital euro will require in practice both
	an existing commercial bank account and a second digital euro account,
	contradicting both easiness and inclusiveness.
We are also not aware of the ECB engaging with experts
	on innumerate users lacking key financial skills.

The ECB's tender
	only requires the reference consumer software
	to be developed for iOS and Android smartphones~\cite{ecbTender009488},
	which are in its vast majority designed and produced outside Europe.
Given that the digital euro will be based on proprietary technology,
	it would only be possible to use the digital euro
	on European mobile phone alternatives
	such as /e/OS and Murena phones or
	Ubuntu Touch from the German UBports Foundation,
	if the ECB actively provided support for it.
At the same time, the digital euro as a proprietary system will prevent its users from
	studying the implementation and challenge its security.
Instead, people will have to trust the provider and the auditors of the ECB
	without possibility to verify their claims.
Truely embracing open standards and free software~\cite{floss} instead of relying on proprietary technology
	would help to establish trust in the system \KF{\Process}
	and allow for easier integration with alternative end-user devices (cf.~\autoref{tab:platforms}).

\begin{table}
	\centering
	\footnotesize

	\begin{tabular}{l c c c c c c c}
	\toprule
	& \multicolumn{5}{c}{Digital Payment System} & \multicolumn{2}{c}{Digital Euro} \\
	\cmidrule(lr){2-6}
	\cmidrule(lr){7-8}
		& Credit Card    & PayPal  & Wero & Bitcoin & GNU Taler & Online  & Offline \\ \midrule
	Android     & \y   & \y   & \y   & \y   & \y    & \y  & \y   \\ \dhline
	iOS         & \y   & \y   & \y   & \y   & \y    & \y  & \y   \\ \dhline
	Browser     & \y   & \y   & \y   & \y   & \y    & \n  & \n   \\ \dhline
	Smartcard   & \y   & \n   & \n   & \y   & \y    & \n  & \y   \\ \dhline
	Terminal    & \n   & \n   & \n   & \y   & \y    & \n  & \n   \\ \midrule
	Open        & \n   & \n   & \n   & \y   & \y    & \n  & \n   \\
	\bottomrule
	\end{tabular}

	\vspace{1em}

	\caption{(Envisioned) platform support of different digital payment systems.
	Despite its two different versions, the digital euro gives less consumer choice.
	The proprietary nature of its design will prevent subsequent expansion of platform support.
	}
	\label{tab:platforms}
\end{table}

\subsection{Would the digital euro replace cash?}
\label{q:cash}

\begin{quote}\scriptsize\sf

No. The digital euro would complement cash, not replace it. The
digital euro would exist alongside cash in response to people’s
growing preference to pay digitally in a fast and secure way. Cash
would continue to be legal tender, and it would remain in
co-existence with the digital euro and any private electronic
means of payment currently in use.

The ongoing euro banknote redesign, with improved security features,
demonstrates that the ECB is committed to the future of cash.
The ECB welcomes the European Commission’s Single Currency Package,
which preserves people’s freedom of choice between cash and digital euro
when paying with central bank money.

\end{quote}

While the ECB may not intend to replace cash with the digital euro,
	the extent to which people and businesses will continue to use physical cash
	once the digital euro is available is not within the ECB's control.
There are several reasons why the digital euro might replace cash at least in practice:
First, since the advent of digital payments,
	every new method introduced has eroded the use of physical cash.
Second, all European businesses will be forced to accept the digital euro.
Third, the costs of operating the system are partially government-subsidized and
	thus likely lower than other digital payment methods.
Fourth, the digital euro (like cash) is backed directly by the central bank.
It seems logical that the digital euro would accelerate the migration away from cash
	more than the introduction of comparable other digital payment methods has done in the past.
The answer by the ECB is clearly downplaying those replacement risks.


\subsection{What value would the digital euro offer merchants?}
\label{q:merchants}

\begin{quote}\scriptsize\sf

The digital euro would be a pan-European solution
allowing merchants to market their products and services to customers all over Europe
with a seamless and consistent payment experience.
It would decrease dependence on non-European payment solutions
by offering a European alternative,
putting merchants in a stronger position to negotiate conditions with payment solution providers,
thereby reducing their own costs.

In developing the digital euro, the ECB is working closely with merchants and their representatives.
The design of the digital euro takes into account what merchants find important,
such as seamless integration with existing checkout systems, ease of use and payment resilience.
The digital euro would also allow merchants to receive payments instantly without additional costs,
even without an internet connection.

\end{quote}

Merchants will have to bear the likely substantial cost
	of integrating support for the digital euro with their business systems.
The legal requirement to accept the digital euro
	at virtually all points of sale across Europe
	will create very high demand for qualified IT integrators,
	which can in turn charge a high premium over typical integration costs.
This will put business owners at a disadvantage
	when negotiating conditions as they will be required
	to roll out support in a market with already short supply.
While the digital euro is planned to be offered free of charge for consumers,
	the current design explicitly includes fees to be paid by merchants
	to payment service providers for access to the digital euro infrastructure \KF{\Economic}.

The proprietary nature of the digital euro technology
	contributes to a challenging integration,
	as integrators will have to work against a ``black box''
	with limited technical details available to them.
Integration is also rarely a one-time expense,
	as the integration needs to be continuously tested and
	maintained as IT systems evolve \KF{\Economic}.

The request for substantial changes to the current digital euro design by merchant associations,
	such as enabling B2B payments and combining the online with the offline version~\cite{merchantscoalition2025},
	clearly shows their disagreement with the current design.
The promised ability to receive digital euros instantly is contradicted
	by the promise made to customers to shop safer online with conditional payments such as pay-on-delivery (cf.~\ref{q:costs-users}).
Receiving digital euro without Internet connection
	is incompatible with the current plan of a zero-euro holding limit for merchants.


\subsection{What value would the digital euro offer payment service providers?}
\label{q:psps}

\begin{quote}\scriptsize\sf

Supervised payment service providers (PSPs), such as banks,
would play a key role in distributing the digital euro.
They would act as the main point of contact for individuals, merchants and businesses
for all digital euro-related matters and would perform all end-user services.

The digital euro could also provide additional business opportunities for PSPs,
giving them an immediate euro area-wide reach.

The ECB’s innovation platform demonstrated the digital euro’s potential
to unify the European payments market and unlock new business models through harmonised standards
and to support future technological developments.
The ECB is using the findings of the innovation platform to inform the further development of the digital euro.

Thus, the digital euro could serve as a platform for PSPs to develop value-added services
within their offer (e.g. conditional payments or loyalty programs).

Moreover, the digital euro compensation model,
as currently envisaged in the European Commission’s proposed digital euro Regulation,
provides PSPs with economic incentives comparable to other digital means of payment.

\end{quote}

The burden of onboarding and support of customer and merchants
	is placed onto payment service providers (PSPs), with no existing cost model \KF{\Economic}.
A recent study among 19 European banks of different sizes, regions and business models
estimates costs of €110 million per bank for introducing the mandatory infrastructure for the digital euro.
This number does not even include costs related to the offline version and merchant acquiring~\cite{pwc2025costs}.

On the other hand, economic incentives for PSPs
	similar to those of other digital means of payments
	beg the question of where cost benefits for the digital euro will come from.

Reasons for PSPs to focus on domestic markets
	have less to do with a fragmention of payment systems
	in the common European market---%
after all, citizens of the eurozone can already use the euro electronically
	and as cash since at least 2002 in the whole area,
	introduced for exactly that purpose.
Rather, differences in legal frameworks
	and regulatory requirements within the euro area,
	e.g. regarding KYC and KYB requirements,
	are holding them back from reaching further than the domestic market.
The introduction of a digital euro
	will not automatically change such fragmented regulatory and legal frameworks.

\subsection{How would the digital euro work?}
\label{q:offline}

\begin{quote}\scriptsize\sf

The digital euro would allow people to make secure instant payments in
physical and online shops and between individuals, irrespective of
the euro area country they are in or which payment service provider they use.

The first step would be to set up your digital euro wallet through
your bank, a post office or other payment service provider.

Once your digital euro wallet is set up, you will be able to put money into it
via a linked bank account or by depositing cash. You would then be able to
make payments using the digital euro wallet, for example via your phone or a smart card.

Digital euro payments would always be safe and instant – whether in
physical stores, in online shops or between people.

The digital euro would offer both online and offline functionalities,
meaning you could use it even when you have poor or no network
reception. Moreover, personal transaction details of offline digital
euro payments would be known only to the payer and the payee,
providing a cash-like level of privacy.

\end{quote}

Ruling out double-spending risks for transactions in offline scenarios
	is impossible  \KF{\Offline} according to the mathematically proven CAP theorem
	which states that no distributed system can be
	partition-tolerant (being disconnected), available (still work), and consistent (without double-spending) at the same time.~\cite{cap}.
An ECB official at a forum in Vienna in 2024 publicly declared this to
	not be a problem since ``secret proprietary technology solves it''~\cite{markpersonal}.
This approach to ``security through obscurity'' contradicts Kerckhoffs' principle---%
	a fundamental concept in cryptography which asserts that
	the security of a system should not depend on the secrecy of its algorithm~\cite{kerckhoffs1883}.
An actual solution would require finding a flaw in the proof of the CAP theorem,
	which must be considered highly unlikely at this point.
In fact, the publicly available documents
	on the current state of the design
	of the offline functionality~\cite{ecbDEoffline2024}
	are still very vague about the actual technical design.
But they do mention the need for ``forgery check during defunding``,
	i.e., a delayed check for double-spending,
	contradicting the claim of offline payments being ``safe and instant'',
	and putting the promised anonymity of offline transactions into question.

The ECB thus acknowledges
	the inevitable risk of double-spending in offline payments,
	hoping to mitigate it with ``secure hardware'' and ``tamper-resistant features'' to
	``protect the information stored on the device
	and allow mutual device authenticity checks''~\cite{ecbDEoffline2024}.
However, focusing on dedicated devices like secure elements (SEs) and eSIMs
	and excluding Trusted Execution Environments (TEEs),
	as recently declared by the ECB in \cite[p.~16]{ecbProgPrep2025},
	will not be enough:
While physical attacks are typically needed to compromise such hardware (cf.~\autoref{fig:offline}),
	the potential return by unlimited digital euro duplication
	will likely outweigh the necessary investment costs.
Offline digital euro wallet holders themselves
	have by definition full physical control of their hardware,
	and therefore must be considered as potential attackers.
That is, any owner of a wallet
	who can acquire the ability (by skill or as service) to break the protection,
	can, in consequence, double spend during offline payments \KF{\Offline}.
The historic track record of breakages of ``tamper-resistant features'' in the past,
	the strong incentive for wallet owners to take advantage of their devices for financial gain,
	and the fair assumption of a large number of deployments of wallets---%
	and therefore opportunities to experiment with the devices in private labs---,
	suggest that such acquisition of ability to double spend is quite likely~\cite{onlinefirst2021}.

With this threat model in mind,
	the selected solution for the offline digital euro
	by G+D~\cite{gdOffline2024,ecb2025tenderawards}
	becomes questionable:
the solution was ranked as ``low'' both for ``Platform Security''
	and ``Maintenance and Communication''
	in a recent survey~\cite{chavanette2024}.
The problem here is not the specific solution,
	but that secrecy
	(i.e. the non-disclosure of the details of the inner workings, in violation of Kerckhoff's principle)
	is the main line of defense against double-spending attacks on digital offline payment systems
	which results in the low ranking in terms of security and transparency for any such platform \KF{\Offline}.

Even without malicious intent of a user,
	the hardware and software may simply just have defects
	that manifest themselves as double-spending or look like fraud.
If those defects are widespread,
	e.g., across many users with the same hardware or software version,
	they would result in many failing attempts of settlement,
	\textit{after} ``successful'' offline payments.
It is worth noting that the damages in these cases
	scale linearly with the number of \textit{transactions} in the system,
	not only with the amount of money in it.
The ECB has not addressed the question of liability,
	i.e., how merchants and users who fall victim to fraud
	of the offline digital euro will be compensated~\KF{\Legal}.
It barely states that
	``either the PSP, the merchant or, in some cases, the consumer would be liable'',
	never the ECB itself~\cite{ecbFinalInv2023}.

Linking a digital euro account to a commercial bank account
	for automatic defunding and funding, in particular, opens up a significant security risk:
A compromised digital euro account could be used as an entry point to the ordinary bank account,
	by moving money without further user confirmation to the digital euro account using the reverse waterfall mechanism,
	thus enabling to steal money from a users' ordinary bank account---%
a threat that didn't exist prior to the introduction of digital euro accounts,
	which are even mandatory in the online version.


\subsection{Who would be able to use the digital euro?}
\label{q:userbase}

\begin{quote}\scriptsize\sf

As stated in the proposed digital euro Regulation presented by the European
Commission, the digital euro would be made available to people,
businesses and public entities that reside or are established in a
euro area Member State on a temporary or permanent basis.

People who travel to the euro area for personal or professional
purposes may also have access to the digital euro.

Moreover, people, businesses and public entities residing or
established outside the euro area may access the digital euro by
opening digital euro accounts with payment service providers
established or operating in a country which is a member of the
European Economic Area or in a third country, subject to a prior
agreement concluded between the EU and the third country concerned and/or
arrangements concluded between the European Central Bank and
the national central bank of the non-euro area Member State or third
country.

\end{quote}

The digital euro is far from being as widely accessible as existing central bank money:
	The online version is inherently tied to accounts at European PSPs
	and onboarding for the offline version entails user identification, too---%
	e.g., to ensure individual holding limits.
These are barriers to access for a significant portion of the European population,
	such as children, undocumented persons, homeless individuals, and others in vulnerable situations.
Physical cash, on the other hand, does not require user identification
	and is independent of complex technologies and costly onboarding.
While it seems likely that no digital payment can ever be as accessible as cash,
	key design choices made by the ECB for the digital euro,
	in particular the digital euro accounts with holding limits per person,
	make the digital euro less accessible than it could be (for example, using digital payment tokens).

Furthermore, the current proposals for the digital euro are
	unclear on how tourists would be on-boarded
	without effectively enabling the digital euro to be used globally and
	potentially becoming a viable alternative to cash for organized crime,
	especially given the transitive anonymous offline function.

\subsection{How private would the digital euro be?}
\label{q:privacy}

\begin{quote}\scriptsize\sf

Privacy is one of the most important design features of the digital
euro.

The digital euro is designed to be able to function offline in a way
that would offer users a cash-like level of privacy, both for sending
money to other people and for making payments in shops. When paying offline,
only the payer and the payee would know the personal transaction
details of the payments made.
Anti-money laundering checks would be carried out by the distributing payment service provider (PSP)
during the funding and defunding process, just as it is the case with cash withdrawals and deposits today.

In the case of online transactions, the Eurosystem would not identify users making or receiving payments,
thereby protecting their personal data, but PSPs would be able to identify users for the purpose of compliance with anti-money laundering rules.

Thus, whether online or offline,
the Eurosystem would not be able to directly link digital euro transactions to specific individuals.

The digital euro would be governed by EU regulations designed to balance
privacy with security. This approach maintains robust protections against
illicit activities, while safeguarding individual privacy.

\end{quote}

Recent surveys commissioned by the ECB show a clear consumer preference for payment privacy:
In the public consultation on the digital euro from 2021,
	more than a third of the respondents ranked payment privacy
	as the most important feature of the digital euro~\cite{ecb2021survey}.
In a recent study about consumer payment habits from December 2024,
	42\% of the respondents affirmed
	that privacy is one of the top advantages of physical cash~\cite{paymenthabits2024}.
But if privacy really was ``one of the most important design features''
	of the digital euro,
	the current design clearly would have failed this promise:
Privacy is only promised for the offline version,
	which is unlikely to work out in practice
	(cf.~\ref{q:offline}),
	while the online version has no privacy advantage
	over existing digital payment systems.

The ECB's recent progress report and other publications~\cite{ecbBlog2024,ecbProgPrepFirst2024,ecbProgPrep2025}
	only reference data \emph{pseu\-do\-ny\-mity} when handled by the ECB
	and compliance with data protection regulations for PSPs,
	but no provable, strong consumer privacy for online transactions.
In fact, the online digital euro offers even \emph{less} privacy to consumers
	than contemporary digital payment methods
	as it centralizes and standardizes the collection of all of its payment data
	for the settlement service at the ECB \KF{\Privacy}.
No effective technical protections are in place
	to prevent the ECB or PSPs from creating ``patterns of life''
	providing detailed insights into citizen's private lives.
As they correctly state, they will ``not be able to directly link digital euro transactions to specific individuals'',
	but could certainly link transactions \emph{indirectly}~\cite{reidentifiability2015},
	and stating that ``the Eurosystem would not identify users'' is just a promise, no guarantee.

To see why pseudonymity is not sufficient for privacy,
	consider the individual transaction histories of persons,
	known to the payment system only by pseudonyms.
The transaction histories span multiple points in time and location,
	including purchases in online stores.
An adversary who also has access to geo-location information of people and their identities,
	e.g., from smartphone vendors and mobile telecommunication providers,
	can now correlate the transaction history
	with geo-location information of individuals,
	and thereby de-anonymize any pseudonym in the transaction history,
	including online purchases.
The scale to which this can be performed
	and the ability to do this basically instantly and permanently,
	makes pseudonymity in a CBDC an enabler of an unprecedented level of easy mass surveillance,
	as payment data is no longer siloed across thousands of organizations,
	databases and incompatible formats~\cite{suerf2022aligny}.

Given the convenience functions of the online version
	such as automatic deposit and withdrawal,
	and given that Internet is available in many situations,
	people are likely to stick to the online form of the digital euro
	and not make use of the offline version at all---%
	probably in the illusion that their transaction data is private there, too \KF{\Privacy}.
The extensive need to identify each individual
	using any digital euro wallet
	hardly compares to the privacy of cash,
	as today we can easily buy wallets for physical cash
	with physical cash without disclosing any personal information.


\subsection{How would the ECB ensure that the digital euro is inclusive and accessible?}
\label{q:inclusive}

\begin{quote}\scriptsize\sf

The digital euro would be a public good, like banknotes and coins
are today, but in digital form.
The ECB is designing the digital euro and the digital euro app
with inclusion as a guiding principle
to ensure users are able to make payments under all circumstances.
The digital euro app would comply with the European Accessibility Act
with a focus on cognitive accessibility to ensure that everyone can quickly learn how to use it.

Informed by user research, as well as interactions with civil society and consumer advocacy organisations,
the digital euro’s design embraces the needs of vulnerable consumers.
Organisations highlighted the importance of a universally accessible solution, intuitive design and in-person support.
Free access to basic digital euro services would also be available to people without a bank account,
closing the digital exclusion gap faced by individuals with no fixed address or beneficiaries of international protection.

Under the proposed digital euro Regulation presented by the European Commission,
banks distributing the digital euro would be required to provide basic digital euro payment services for free when requested by their customers.

The digital euro would be designed to accommodate the needs of
everyone, leaving no one behind.

\end{quote}

Physical cash is the most inclusive form of payment,
	and the digital euro will likely be detrimental to cash use (cf.~\ref{q:cash}).
Holding and paying with digital euros
	will require ownership of supported hardware,
	and the skills to use it.
Today, many citizens struggle with modern banking applications
	requiring multi-factor authentication,
	and we cannot find any answers for this challenge
	in the published designs for the digital euro.
In 2024, between 7\% and 8\% of citizens in the euro area, across all age groups, had no access
	to an account or a card for digital payments, respectively~\cite{paymenthabits2024}.
Moreover, the ECB has given no concrete indications
	as to how it plans to work on solutions for illiterate or innumerate users.
It is also left unclear who will pay for the additional workload and costs
	connected to face-to-face services by public entities or participating PSPs
	so that ``no one is left behind''.
Overall, the central bank underestimates the causes of the digital divide
	and the digital euro will likely deepen the rift between those
	with access to modern technology and those living without.

\subsection{How would the ECB ensure that digital euro payments work in the same way throughout the euro area?}
\label{\thesubsection}2026

\begin{quote}\scriptsize\sf

Supervised payment service providers (PSPs), such as euro area banks,
would be responsible for distributing the digital euro.
To ensure the digital euro scheme is implemented in the same way across the entire euro area,
the Eurosystem is developing a digital euro scheme rule book
in a collaborative and iterative process with market participants.
The rulebook would establish a single set of rules, standards and procedures
to ensure consistent basic digital euro services throughout the euro area,
providing a uniform experience for users
regardless of the Member State they are in or the PSP involved – as is the case with cash today.

\end{quote}

Instead of using common terminology from engineering and computer science,
	and promoting well-defined \emph{specifications}
	for \emph{protocols} for the interaction of different participants in the system,
	the ECB uses terminology from the legal field of compliance (``rulebook''),
	thus indicating that the focus of the ECB lies in legal aspects of the CBDC,
	not the actual engineering challenges.
Also, setting narrow rules, standards and procedures
	stands in conflict to the claim
	of the digital euro fostering innovation,
	and real customer choice.
It rather promotes a financial monoculture
	where failures are more likely to result in wide-spread outages,
	similar to the recent CrowdStike outage
	in the Western world standardized on Microsoft technology~\cite{crowdstrike}.

\subsection{Would the digital euro be an alternative currency within the Eurosystem?}
\label{\thesubsection}

\begin{quote}\scriptsize\sf

No. Just as banknotes and coins are not alternative currencies,
but rather different forms of the same currency,
the digital euro would be just another way to pay in euro.
The digital euro would accommodate people’s and firms’ growing preference to pay digitally.

\end{quote}

Actually, the digital euro is not just ``another way to pay'',
	but two distinct new methods of payment
	with very different properties.
Relating it to banknotes instead of bank accounts
	may create the illusion that both the online and the offline version are comparable to cash,
	which is only the case insofar that both are issued by the ECB.
Crucial differences with respect to privacy are further discussed in \ref{q:privacy}.

While there is a general trend towards digital payments both online and in shops,
	reading that as a ``preference'' may be oversimplifying,
	especially as digital payment methods come with associated costs that
	are often not transparent to the buyer as those are absorbed in the price,
	independent of the specific cost structure of the payment method.
There are also instances of firms being coerced
	into offering and using digital payment methods,
	which should not be mistaken for a chosen preference:
For example, Italy was mandated by the European Union
	to require shops to accept credit card payments,
	even for small transactions~\cite{italy2022cardforce}.
At the same time, some shops refuse to accept cash,
	forcing customers to pay with cards~\cite{letemps2020refusDeCash}.
Mastercard has taken advantage of the COVID-19 pandemic
	to run false campaigns about ``unhealthy cash''
	in order to promote its business interests~\cite{mastercard}.
As an ECB study on the payments attitudes of consumers rightfully mentions,
	the decline in cash usage in shops from 72\% (2019) to 59\% (2022)
	was heavily influenced by the pandemic,
	and ``may prove temporary'' \cite{paymenthabits2022}.

\subsection{What would be the link between instant payments and the digital euro?}
\label{\thesubsection}

\begin{quote}\scriptsize\sf

Today, when consumers make cashless payments in shops, merchants don’t receive
the money immediately. The digital euro would change that – all
digital euro payments would be instant.

The single set of rules, standards and procedures being developed
for the digital euro would mean that instant
payment solutions could be further developed to reach all euro area
countries. This would reduce Europe’s dependence on the small number of private non-European
companies that currently dominate the payments sector.

\end{quote}

While today money spent at shops
	is indeed most of the time not immediately transferred to the merchant's account,
	there do already exist more modern instant settlement methods for payments in shops
	(such as Twint in Switzerland, Swish in Sweden, or Vipps in Norway,
	which incidentally are all operated by European companies)
	and---more prominently---Instant SEPA transfers within the euro area.
The digital euro in its account-based online version
	has no technical advantage over these other options
	for citizens residing in a single country.

A key objective of the digital euro seems to be to limit market access
	for non-European instant payment service providers, limiting
	consumer choices. This objective seems incompatible with
	Article 119 of the Treaty for the Functioning of the European Union (TFEU)
	establishing the principle of an open market economy with free competition~\cite{guetschow2025deregulation}.

On the other hand,
	payments with the offline version of the digital euro
	will need Internet connectivity to be finally cleared, i.e., are not instant \KF{\Offline}.
In case of failures, it is unclear who bears the liability,
	or when the failure would even be detected \KF{\Legal}.

\subsection{How would the digital euro’s technical architecture work, and would it be based on distributed ledger technology such as blockchain?}
\label{\thesubsection}

\begin{quote}\scriptsize\sf

The digital euro would operate on a centralised settlement platform
and the Eurosystem would record and verify all settlements and holdings.
As direct liabilities of the Eurosystem,
it is important that the digital euro in people’s wallets are safe in order to maintain trust,
both in the euro and in the Eurosystem.

The digital euro is not based on distributed ledger technology (DLT),
but it makes use of key design principles from DLT to enhance resilience and efficiency
and to improve the system’s overall performance and reliability.

The digital euro’s resilient technical architecture would be built on established standards.
A multi-region setup in which each region is equipped with multiple servers,
going well beyond standard redundancy models, will ensure service continuity under all circumstances.

\end{quote}

It is positive that the ECB finally acknowledged in the most recent revision of the FAQ
that technology created to {\em decentralize} payments is not the best fit for a {\em central} bank.
Until October 2025, it still stated that technical decisions were yet to be taken~\cite{defaq2025old},
although central settlement at the ECB was already part of the first proposal in 2020~\cite{ecb2020}.
Maybe it was politically difficult to simply reject DLT,
after having hired blockchain experts~\cite{ezb2024duve},
and while being subject to lobbying from cryptocurrency businesses~\cite{dea2024members}.

\subsection{Where does the digital euro project currently stand?}
\label{q:status}

\begin{quote}\scriptsize\sf

The digital euro project is moving forward.
The preparation phase, launched in November 2023 and concluded in October 2025,
advanced technical development and learning through experimentation.
This work built on the design choices and technical requirements defined during the investigation phase.

Following on from the investigation and preparation phases,
the Eurosystem is now making further progress on technical work,
deepening market engagement and continuing to support the legislative process.
We aim to be ready for a potential first issuance of the digital euro during 2029,
assuming the necessary regulation on the establishment of the digital euro is adopted in the course of 2026.

\end{quote}

The digital euro is moving forward very slowly.
Wero, a cross-border payment system initiated by the private European Payments Initiative,
has evolved from first announcement in September 2023
to 40 million registered users in four European countries by the end of 2025,
aiming to include more countries and to offer retail payments in 2026~\cite{epi2025wero}.
The digital euro project, on the other hand,
has been ongoing for more than five years,
starting with a first report by the ECB in 2020~\cite{ecb2020}.
In case the first digital euro will be issued in 2029 as promised,
private competitors might have developed far enough
so that one genuine advantage of the digital euro,
euro-area-wide availability of a European payment system,
is no longer a distinguishing factor.

Fernando Navarette, the European Parliament's rapporteur on the digital euro regulation draft,
has raised similar concerns~\cite{navarette2025position}
and recently proposed changes to the legislative framework
which would only introduce the online version of the digital euro
on the condition that no private pan-European payment system exists~\cite{navarette2025report}.
If a private solution exists, only the offline version of the digital euro would remain,
with all the risks discussed in \ref{q:offline}.

\subsection{What is the digital euro pilot?}
\label{q:pilot}

\begin{quote}\scriptsize\sf

The digital euro pilot is an exercise organised by the Eurosystem
to assess how the digital euro could work in practice.
Participants will use a beta version of the digital euro
in real-life situations to validate its technical functionalities,
operational processes and user experience.
The pilot is planned to start in the second half of 2027 and will run for 12 months.

During the pilot, staff from participating Eurosystem central banks
will make everyday payments using the beta digital euro.
For example, they will be able to send money to each other (both online and offline)
and pay selected merchants, such as cafeterias, restaurants
and online shops linked to the ECB and national central banks.
The beta digital euro will be used only for the pilot
and is intended to work in a similar way to a possible future digital euro.

The participating PSPs, such as banks, will be selected by the Eurosystem
and will provide the services necessary to support these payments.

What we learn from the pilot will support the Eurosystem’s ongoing preparatory work
and help inform future decisions on the digital euro to ensure
that, if the digital euro is introduced, it is fully equipped to meet the needs of all Europeans.

\end{quote}

The digital euro pilot needs to be restricted to a clearly defined user group
to not fall under existing regulation of digital payment systems.
It is natural to restrict the user group to Eurosystem staff,
who will be able to assess the normal operation of the system.

However, this target group will most probably not have enough technical expertise
to perform thorough security testing.
If no external researchers, security experts or hackers are given access to the pilot,
security issues such as double-spending in the offline version \KF{\Offline}
will only surface once the digital euro is introduced and available to the public.

\subsection{Who is involved in the digital euro project?}
\label{q:consultation}

\begin{quote}\scriptsize\sf

The Eurosystem – which comprises the ECB and the national central banks of
the euro area – aims to ensure that the digital euro meets users’ needs. For
this reason, the Eurosystem engages regularly
with policymakers, legislators, market participants, civil society
organisations and members of the public, who would be the ultimate users of the digital euro.

This engagement takes place in various contexts, such as the
Euro Retail Payments Board, which
brings together stakeholders from all parts
of the European retail payments market, and the Rulebook
Development Group, which comprises senior professionals
from the public and private sectors with experience in finance and payments (see Q18).

The ECB also regularly engages with:
\begin{itemize}
	\item private companies, which provide feedback on the technical
	aspects of the digital euro based on their market knowledge and
	testing on the digital euro innovation platform;
	\item European civil society organisations via seminars
	to listen to their views and foster an open dialogue;
	\item potential end users through surveys, interviews and
	focus groups to understand their needs and preferences.
\end{itemize}

The ECB regularly participates in Eurogroup meetings with the finance ministers
of euro area countries and provides regular updates on the digital euro project
to the Committee on Economic and Monetary Affairs of the European Parliament.

\end{quote}

The public consumer consultation~\cite{ecb2021survey} showed a clear preference for payment privacy
	and, with less importance, the wish for payments without internet connectivity.
While practical proposals for a privacy-preserving CBDC design do exist~\cite{suerf2021moser,suerf2022aligny,platypus,uhlig2023privacy},
	the digital euro does not offer any privacy advantages in its online version (cf.~\ref{q:privacy}).
Preventing double-spending without Internet connectivity, on the other hand,
	which is promised for the offline version of the digital euro,
	contradicts mathematical evidence \KF{\Offline} as further detailed in \ref{q:offline}.

It is questionable to which extent the ECB was open
	to external suggestions during the design process,
	given that most key aspects of the current design---%
	such as the separation of the online and offline version,
	the waterfall approach of linking commercial bank accounts to digital euro accounts,
	the two-tiered architecture involving private PSPs,
	or the usage of ``secure hardware'' for the offline version---%
	were already proposed by the ECB
    \textit{before} the start of the public consultation phase
    \KF{\Process}~\cite{digitaleuro2020}.

The ECB engagement with private companies has excluded small and medium enterprises
	by setting high thresholds for potential participants
	already during the investigation phase \KF{\Process}:
``Candidates must meet the following minimum requirements:
	a) the average annual total net turnover of the Candidate
	must be at least EUR 100,000,000
	for the last three financial years; and
	b) the average annual net turnover
	of the similar services covered
	by the contract must be at least EUR 10,000,000
	for the last three financial years.''~\cite{ecbTender0078480}
Requirements for tender application in the preparation phase were similar~\cite{ecbTender009488}.

Organizing seminars to ``listen'' to the European civil society
\emph{after} making key design decisions is not an honest invite in participation,
but a public relations exercise \KF{\Process}.
An earlier open dialog with stakeholders and experts
with the objective of identifying a feasible design for the digital euro
that would have social benefits beyond establishing ``absolute control''~\cite{bis2021absolute}
over the population by the central bank would have been preferable~\KF{\Benefits}.

\subsection{How are European legislators involved in the process?}
\label{q:legislation}

\begin{quote}\scriptsize\sf

On 28 June 2023 the European Commission presented the Single Currency Package
containing proposals to support the use of cash and to establish the framework for a possible digital euro.
The ECB welcomes the fact that the digital euro proposal is accompanied
by a proposal to strengthen the role of cash,
as both would be legal tender and forms of central bank money.
The purpose of the proposed digital euro Regulation is to ensure that any future digital euro would give
people and businesses the option to pay digitally using a
widely accepted, cheap, secure and resilient form of public money
anywhere in the euro area.

The ECB provides support and technical input during the legislative process, as required.
The Eurosystem will consider any necessary adjustments to the design of the digital euro that may emerge from legislative deliberations.

The ECB’s Governing Council will not make a decision on
whether to issue the digital euro until the Regulation on the establishment of the digital euro has been adopted.

\end{quote}

While the purpose of the legislation should indeed be
	to provide a cheap, secure and resilient form of public money,
	the digital euro project conducted by the ECB fails to meet these expectations:

First, the amount of €1.3 billion spent on the project (cf. \ref{q:costs-system})
	contradicts	the notion of a ``cheap'' form of public money,
	given that other modern payment systems have been created
	with budgets of less than 1\% of that amount~\cite{tsys}.
Also note that this amount is only the budget
	for developing the digital euro at the ECB and
	excludes the much higher cost of actually introducing the system across the euro area \KF{\Economic}.
Second, the design proposed by the ECB
	suffers from several inherent security flaws,
	such as the ``reverse waterfall'' exposing money in associated commercial bank accounts
	to risks from compromised digital euro accounts,
	or the offline functionality contradicting Kerckhoffs' principle \KF{\Offline} as discussed further in \ref{q:offline}.
Third, given the additional complexity on top of existing payment infrastructure for the online version \KF{\Benefits},
	the proclaimed improved resilience can only be achieved with the offline functionality
	and is commented on in more detail in \ref{q:resilience}.

Running (legislative) design and implementation processes in parallel
	is a recipe for cost explosions,
	likely pushing the cost beyond the budgeted €1.3 billion.
The ECB is spending a large sum on the project prior
	to having a legal mandate for it,
	thus creating a \textit{fait accompli} bypassing the democratic process and exceeding its mandate.

\subsection{How is the digital euro scheme rulebook being developed?}
\label{q:rulebook}

\begin{quote}\scriptsize\sf

The Eurosystem is developing the draft rulebook in close collaboration
with representatives of the European retail payments market through the Rulebook Development Group (RDG).

The RDG, which consists of senior representatives from European associations
representing both the supply and demand sides of the European retail payments market,
is working on the basis of the digital euro design choices that have already been approved by the ECB’s Governing Council.

Dedicated workstreams have been created within the Rulebook Development Group
to focus on sections of the rulebook that require particular expertise.

In June 2025 a revised interim draft of the digital euro scheme rulebook was delivered to the RDG for a market consultation.
The draft rulebook remains sufficiently flexible to accommodate any future adjustment
deriving from the final text of the Regulation on the establishment of the digital euro, once adopted.

\end{quote}

Instead of engaging in public discussion
	with the information security community,
	the rulebooks are developed behind closed doors \KF{\Process}.
The digital euro design choices cannot be questioned
	by the experts working on the rulebooks,
	regardless on the privacy, security or
	usability issues they may create.
The digital euro rulebook should better be seen as an attempt to use red tape
	to paper over the inherent design problems of the digital euro
	by prescribing complex processes and
	imposing costly business process requirements on participants \KF{\Economic}.
Instead of developing innovative technical standards,
	it will likely primarily refer to established norms and procedures.

\subsection{Would the digital euro be programmable money?}
\label{\thesubsection}

\begin{quote}\scriptsize\sf

Programmable money is a digital form of money used for a predefined purpose,
like a voucher, with limitations on where, when or with whom it can be used.

As also envisaged in the European Commission’s proposed digital euro Regulation,
the digital euro would never be programmable money,
but it could facilitate conditional payments
(for example, if a customer buys something online and chooses the option to pay on delivery).

\end{quote}

It is strange that the ECB would use such an absolute term as ``never''
	in the context of a technology that can easily be modified or extended based on sociopolitical demands,
	especially given that the ECB is allegedly still under public consultation with stakeholders.
The answer also does not provide the reasons
	why the ECB has a different view on programmable money than other central banks
	such as the Monetary Authority of Singapore~\cite{masProgrammable2023}.
There are valid use cases for such technology,
	some of which are politically demanded and arguably socially beneficial,
	like privacy-preserving mechanisms for age restriction~\cite{kesim2022zero}.
Such features should therefore probably be considered on a case-by-case basis.

It is not unlikely that private alternatives will offer digital payments with programmable money
even before the issuance of the first digital euro,
leaving the ``global front-runner'' by the ECB behind.

\subsection{Would people have to pay to use the digital euro?}
\label{q:costs-users}

\begin{quote}\scriptsize\sf

As a public good, the digital euro would be free for basic use by individual users.

Banks or other payment service providers could offer
their customers additional, paid digital euro services.
Such value-added services could make the digital euro even more attractive to users,
for example for making conditional payments.
These could allow customers to shop safer online,
with money only being transferred when the delivery of the product has been confirmed,
thereby reducing the risk of fraud and simplifying refunds.

\end{quote}

Even the absence of direct costs for end users cannot hide the fact that
	there are significant underlying expenses associated with the implementation of a digital currency.
These include the costs of initial integration and onboarding,
	system maintenance, operational expenses, end-user devices for offline payments,
	and compliance with legal requirements such as Know-Your-Customer (KYC).
The current design envisions costly payment infrastructure and customer support by PSPs (cf.~\ref{q:psps}),
	(taxpayer-funded) public entities to provide end-user onboarding,
	mandatory acceptance at European merchants for a fee (passed on to consumers via higher prices, cf.~\ref{q:merchants}),
	and high expenses for development, maintenance and operation,
	borne but not paid by the ECB (cf.~\ref{q:costs-system}).
Reduced profits for the ECB in turn reduce the profits of the national central banks
	and thereby the dividends paid by central banks to the national budgets,
	which finally leads to higher taxes.
Ultimately, all costs for the digital euro will be borne by the citizens of the euro area.

The given example of an escrow service
	lacks a convincing argument how this is \emph{added value},
	as banks and digital payment providers can provide those services already,
	without the digital euro.
Additionally, it presents an overly simplistic perspective on the intricacies of online fraud
	by proposing a ``solution'' that focuses solely on one interest---the user's---
	while disregarding others. This makes it unlikely to persuade merchants.

\subsection{How do other pan-European payment initiatives relate to the digital euro?}
\label{\thesubsection}

\begin{quote}\scriptsize\sf

The ECB welcomes European market initiatives that reach beyond domestic markets.

The digital euro should enable domestic and regional schemes to scale up
across different use cases and across borders,
facilitating easier, broader and more efficient acceptance of European private sector solutions
thanks to the use of harmonised standards.
European payment service providers stand to benefit from these opportunities,
primarily through increased geographical reach and use cases not previously served.

The design envisages the possibility of integrating private solutions through,
for instance, possible co-badging on physical cards and existing digital wallets.
In both cases the digital euro would be the “fall-back” that enables full pan-European reach
while preserving market access for domestic or regional schemes where they are accepted.

\end{quote}

This question has been added in a recent revision in October 2025~\cite{defaq2025old},
likely as a response to the position paper by the European Parliament's rapporteur
Fernando Navarette~\cite{navarette2025position},
which rightfully points to existing initiatives in the private sector (cf. \ref{q:status}).
While it may sound reassuring that ``the ECB welcomes [such] initiatives'',
it stays unclear why we would need harmonised standards
that go beyond what is already available with Instant SEPA transfers,
and why we could not achieve such standardization without introducing a CBDC.

\subsection{Would payment service providers (PSPs) be compensated for distributing the digital euro?}
\label{q:psps-compensation}

\begin{quote}\scriptsize\sf

The Eurosystem is proposing a compensation model
that would create fair economic incentives for all parties involved in the digital euro ecosystem.
For banks and other PSPs, the compensation model addresses the operational costs of distributing the digital euro.

As is currently the case with other payment systems,
PSPs distributing the digital euro would be able to charge merchants for these services.
Price setting for merchants and PSPs would be subject to a cap,
as proposed by the European Commission in its digital euro Regulation.

As with the production and issuance of banknotes,
the Eurosystem would bear the costs of the establishment of the digital euro scheme and infrastructure.
Moreover, the Eurosystem would aim to minimise additional investment costs for PSPs by reusing existing infrastructures as much as possible.

\end{quote}

Existing large payment service providers with extensive KYC data
	will likely be able to monetize their existing customer relationships
	to profitably offer services related to the digital euro,
	while new and smaller businesses with higher initial costs
	may be effectively excluded by the fee cap
	and the uniformity of the service offering \KF{\Economic}.

Given that the digital euro will be a liability of the central bank,
	payment service providers will not just compete mostly on costs;
	they will be able to do so at the expense of security
	as they are not ultimately liable for the digital euro~\KF{\Legal}.
Thus, we predict that running systems cheaply with minimal regard for security---%
	or even committing outright fraud at the expense of the central bank---%
	will be the competitive drivers among private digital euro payment service providers.

\subsection{Would the digital euro pose a threat to financial stability by disintermediating banks?}
\label{q:financial-stability}

\begin{quote}\scriptsize\sf

Our financial system – with the banking system at its centre – functions well,
and the Eurosystem wants to preserve the key role banks play in ensuring the efficient provision of credit to the economy.

The ECB has made the following design choices to minimise
any potential risks the digital euro might pose to the financial system.

\begin{itemize}
	\item Users would only be able to hold a limited amount of digital euro in their account.
	This would prevent excessive outflows of bank deposits and help preserve the stability of our financial system, even in times of crisis.
	\item Linking their digital euro wallet to a bank account would allow users to make payments above the holding limit
	and cover any shortfall instantly without having to pre-fund their digital euro wallet
	(assuming sufficient funds are available in the linked account).
	\item As with cash in your wallet, no interest would be paid on digital euro holdings.
\end{itemize}

The ECB prepared a technical analysis to estimate the potential effects of various hypothetical holding limits,
following a request that emerged during legislative negotiations.
This analysis confirmed that using the digital euro for day-to-day payments would not harm financial stability
and that – given the different hypothetical holding limits of up to €3,000 per person that the co-legislators asked to be tested –
the impact of the digital euro would not harm financial stability within the euro area,
even under a highly unlikely and extremely conservative crisis scenario.

\end{quote}

In contrast to the analysis done by the ECB,
	independent studies have come to different conclusions
	showing that especially highly impacted small banks would still experience significant threats
	to their financial stability:
Loosing up to 20\% of the deposit base or 9\% of total bank liabilities
	according to~\cite[p.~10]{effects2023},
	or 15\% of total bank liabilities according to~\cite[p.~18]{knowlimits2023}.
While customers could withdraw similar amounts in physical cash today,
	the digital euro will allow them to do so faster and from anywhere in the world,
	limiting the time-frame the banking system has to react to a bank run.

Many banks are only profitable because of revenue from account and payment fees~\cite{deyoung2004banks}.
If gratis digital euro accounts and transactions were to deprive them of this revenue,
	they will be forced to adjust their business models to
	make up for this loss of profits \KF{\Economic}.
One such business model might be that banks could earn interest from the central
	bank on the digital euros under their management via the ECB's deposit
	facility, costing the ECB significant interest on all digital euros in circulation~\cite{guetschow2025deregulation}.

Apart from that,
	the offline functionality of the digital euro
	poses another threat to the financial stability of the euro \KF{\Offline}:
As soon as someone overcomes the double-spending restrictions (cf.~\ref{q:offline})
	they will effectively be able to create digital euros at an unlimited scale.
To prevent devaluation of the (digital) euro in such a case,
	the offline functionality would need to be completely disabled
	(assuming that would be even possible without Internet connection)
	until the cause is fixed, and all affected hardware is replaced.

\subsection{Would the introduction of the digital euro make payments in Europe more vulnerable to cyberattacks?}
\label{q:cyberattacks}

\begin{quote}\scriptsize\sf

As with other digital infrastructures, the digital euro could be a target for cyberattacks.
To mitigate this risk, the design of the digital euro would employ state-of-the-art technologies
to create a cyber-resilient and future-proof environment.
In the design of the cybersecurity controls,
the ECB is making use of proven Eurosystem practices from other market infrastructures
and regular planned testing against simulated attacks.

\end{quote}

Clearly, ``could be'' is an understatement,
	we can be sure that it ``will be'',
	given that the digital euro settlement system would offer a pan-European record of payment data.
The promise of a ``cyber-resilient (...) environment''
	lacks convincing support in published documents~\cite{ecDE2023,ecbFinalInv2023,ecbProgPrep2025}:
	The ECB barely refers to buzzwords such as ``secure development practices'' and ``reduction of the attack surface'',
	without providing further details~\cite[p.~17-18]{ecbProgPrep2025}.
	At the same time, they do mention the usage of ``sophisticated AI models'' in the context of the digital euro,
	which opens a whole new class of attacks vectors.

The formulation of ``state-of-the-art technologies'' is particularly misleading,
	as it does not imply the application
	of the most modern or secure solution available,
	but merely ``what everybody else does''.
For example, Windows with CrowdStrike in the Cloud
	was considered state of the art
	for the Western airline industry as of summer 2024,
	rendering it severely vulnerable
	exactly \textit{because} of the usage of CrowdStrike software
	which is designed to infiltrate all important parts of a system~\cite{crowdstrike}.
The current state of the art is everything,
	but not cyber-resilient or future-proof.

Furthermore, the state of the art is unclear
	in the context of transitively anonymous offline payments \KF{\Offline}.
As the ECB confirms in \cite[p.~16]{ecbProgPrep2025}, there are no other large-scale payment solutions
	that currently support such offline payments, making the claim questionable.

\subsection{How would the digital euro be different from stablecoins and crypto-assets?}
\label{q:crypto}

\begin{quote}\scriptsize\sf

The digital euro would be central bank money, issued and guaranteed by the Eurosystem,
which comprises the European Central Bank and the national central banks of the euro area.
Like euro banknotes and coins, it would be legal tender,
meaning everyone would be able to use it for payments.
As central bank money and a public good, it would be stable and reliable –
you would always be able to trust that one digital euro is worth one euro.

Stablecoins are created by private companies.
They are not guaranteed by a central bank or public authority.
Their value depends on how well the company manages its reserves and finances,
and this can be influenced by factors outside their control.
This means their stability is not as certain as that of the euro.

Crypto-assets such as Bitcoin or Ether are different again.
They are not backed by any institution and have no underlying value.
Their prices can go up and down sharply,
and there is no organisation responsible if they lose their value.

\end{quote}

While it may be justified to expect that a CBDC is more reliable
	compared to one issued by \emph{private} (potentially unregulated) parties,
	framing the digital euro as ``stable and reliable'' is an oversimplification.
Same as any payment scheme, the digital payment system that underpins the digital euro
	is still exposed to a broad range of risks.
These include
	operational risks,
	fraud and security risks,
	liquidity risks,
	credit risks,
	settlement risks,
	technology risks,
	payment system arbitrage risks,
	third-party risks, and
	privacy breach risks~\cite{bis1993}.
The ECB, as the responsible public institution and operator of the digital euro,
	must take these risks into consideration, provide sufficient mitigations,
	and communicate them openly~\KF{\Legal}.

In addition to the mentioned risks,
	the stability and reliability of the euro---%
	not just the digital one---%
	ultimately depend on the actions of the ECB and its credibility,
	which is the basis of public trust in its currency.
Seen in light of this connection of public trust and credibility,
	the decisions of both the Swiss central bank and the U.S. Federal Reserve
	not to issue a retail central bank digital currency,
	are likely in part due to the perceived risks to their reputation~\cite{snbCBDC2024,fedCBDC2024}.

Presumably to ensure transaction privacy and data protection,
	the digital euro is designed as a two-tiered system,
	i.e., with a separation between the core transaction system and
	the customer identification system,
	following precisely the Libra/Diem design~\cite{libra2021} by Meta (formerly Facebook).
However, unlike Meta,
	the government is effectively immune to fines
	imposed for privacy breaches,
	as it would merely pay the fine to itself.

Even worse, enforcing claims against the ECB
	will be harder than enforcing claims against private companies.
The ECB has political independence
	and an effectively unlimited legal budget,
	and can claim sovereign immunity.
Thus, enforcing claims against the ECB
	is likely only to succeed in cases
	which the bank chooses not to actually fight.
In summary, citizens have no effective remedy
	against breaches of the ECB's mandates
	on data protection \KF{\Privacy}.
In contrast, fraudulent cryptocurrency businesses
	have been successfully prosecuted
	and victims have received compensation~\cite{eurojustCrypto2024}.

Finally, there is a large difference
	between the private and the public sector
	with respect to what personal data enables them to do.
Public authorities can change the rules and allow later something
	that is prohibited at a given time.
They could for instance legalize the use of data for criminal and political persecution.

\subsection{How much would the digital euro project cost the Eurosystem?}
\label{q:costs-system}

\begin{quote}\scriptsize\sf

Investing in the digital euro is key to ensuring our currency and
payments sector remains fit-for-purpose in the digital age.

Some of the digital euro components, such as payment settlement,
would be developed internally within the Eurosystem.
For others, like the offline services component,
we have established framework agreements with external providers.
Framework agreements do not involve any payment and include safeguards allowing the scope to be adjusted in line with changes in the legislation.

Total development costs,
comprising both externally and internally developed components are estimated to amount to €1.3 billion,
while annual operating costs are projected to be around €320 million.
The Eurosystem is continuing preparations in response to calls from euro area leaders
to be ready for potential issuance as soon as possible.
However, the necessary legislation has not yet been adopted.
The work is therefore structured in modules to allow gradual scaling and to limit financial commitments.

The Eurosystem would bear the costs of the establishment of the digital euro scheme and infrastructure,
just as it does for the production and issuance of euro banknotes – which, like the digital euro, are a public good.
As with banknotes, these costs would be covered by “seigniorage” (the income the ECB earns from issuing money)
even if digital euro holdings were small compared with banknotes in circulation.
The ECB is committed to keeping costs low by reusing existing infrastructure as much as possible,
while still delivering a digital euro that brings value to consumers and merchants.

In line with its nature as a public good,
the digital euro would be free for basic use for consumers and cost efficient for European merchants.
The Eurosystem would not charge or benefit from any digital euro transaction fees.

\end{quote}


The framework agreements sum up to €1.164 billion~\cite{ecb2025tenderawards},
leaving less than €250 million for in-house development of the settlement infrastructure---%
despite it being the apparently more complex heart of the system~\cite[Fig.~4]{ecbProgPrep2025}.
Compared with the costs of the external components, this does not seem a lot.
Contrary to its proclaimed goal of keeping costs low by re-using existing infrastructure,
	the ECB has not considered modern and already developed digital payment systems \KF{\Process},
	such as GNU Taler---which is free software and has been created with less than 1\% of the public tenders' amount~\cite{tsys}.

The answer largely passes over the high cost of initial integration and continuous support
	faced by merchants throughout the whole euro area to comply with the legally mandated acceptance of the digital euro (cf.~\ref{q:merchants}).
It also forgets to mention costs to remedy anticipated fraud in the offline version (cf.~\ref{q:offline}),
	(the PSPs') typical operational losses due to illicit access to digital euro accounts in the online version,
	or compensation for data breaches from the ECB's central database of all digital euro transactions \KF{\Privacy}.
Finally, it ignores the costs of changing design requirements from the legislation process
	that is still ongoing in parallel to the implementation (cf.~\ref{q:legislation}),
	and the possibility of a complete project failure where all invested money would be a loss.

According to the proposed regulation,
	``the digital euro should be issued (...) by converting payment service providers'
	central bank reserves into digital euro holdings''~\cite[Rec. 9]{ecDE2023}.
Hence, no seigniorage income will be generated since the central bank will not acquire any new assets by issuing digital euros.

}

\end{document}